\documentclass[12pt]{iopart}
\usepackage{geometry,graphicx, iopams,float,subfigure,subcaption,xparse,pgffor}

\expandafter\let\csname equation*\endcsname\relax

\expandafter\let\csname
 endequation*\endcsname\relax
\usepackage{amsmath}
\usepackage{lineno}
\graphicspath{{./figures/}}

\newcommand{\figref}[1]{Fig. \ref{fig:#1}}

\begin{document}

\title{Integrals and chaos in generalized H\'{e}non-Heiles Hamiltonians}
\author{G. Contopoulos$^{1}$, A.C. Tzemos$^{1}$\footnote{Corresponding Author} and F. Zanias$^{1,2}$}

\address{$^1$Research Center for Astronomy and 
Applied Mathematics of the Academy of 
Athens - Soranou Efessiou 4, GR-11527 Athens, Greece}
\address{$^2$ University of Amsterdam, Science Park 904, 1098 XH Amsterdam, The Netherlands}

\ead{gcontop@academyofathens.gr,atzemos@academyofathens.gr,\\foivos.zanias@student.uva.nl}
\vspace{10pt}

\begin{abstract}
\par We study the approximate (formal) integrals of motion in the Hamiltonian $ H = \frac{1}{2}\left( \dot{x}^2 + \dot{y}^2 + x^2 + y^2 \right) + \epsilon\,\left( xy^2 + \alpha x^3\right)$ which is an extension of the usual H\'{e}non-Heiles Hamiltonian that has $\alpha = -1/3$. We compare the theoretical surfaces of section (at $y=0$) with the exact surfaces of section calculated by integrating numerically many orbits. For small $\epsilon$, the invariant curves of the theoretical and the exact surfaces of section are close to each other, but for large $\epsilon$ there are differences. The most important is the appearance of chaos in the exact case, which becomes dominant as $\epsilon$ approaches the escape perturbation for $\alpha<0$. We study in particular the cases $\alpha = 1/3$, which represents an integrable system, and $\alpha = 0$. 
Finally we examine the generation of chaos through the resonance overlap mechanism in the case $\alpha=-1/3$ (the original H\'{e}non-Heiles system) by showing both the homoclinic and the heteroclinic intersection of the asymptotic curves of the unstable periodic orbits.
\end{abstract}

\section{Introduction}\label{sec1}

The ``third integral'' of motion was developed many years ago
in axisymmetric galactic models \cite{Contopoulos1960,Contopoulos1963}(early formulations were made by Whittaker \cite{Whittaker1916,Whittaker1937} and Cherry \cite{Cherry1924a,Cherry1924b}. These systems have a first integral which is the total energy (sum of potential and kinetic energy), and a second integral, the axial component of the angular momentum. The ``third integral'' refers to a potential in the variables $x$ and $y$ ($x$ on the plane perpendicular to the axis  of symmetry and $y$ along the axis) around the center of a galaxy. A usual form of this Hamiltonian is 

\begin{equation}
H=H_0+\epsilon H_1+\epsilon^2 H_2+\dots,\label{hams}
\end{equation}
where 
\begin{equation}
H_0=\frac{1}{2}(\dot{x}^2+\dot{y}^2+\omega_x^2x^2+\omega_y^2y^2)\label{H0}
\end{equation}
and $H_n$ is a polynomial of degree $n+2$ in $x$ and $y$.

The third integral is a series of the form
\begin{equation}
\Phi_i=\Phi_{i,0}+\epsilon\Phi_{i,1}+\epsilon^2\Phi_{i,2}+\dots,\quad (i=1,2)\label{third}
\end{equation}
where
\begin{equation}
\Phi_{1,0}=\frac{1}{2}(\dot{x}^2+\omega_1^2x^2)
\end{equation}
or
\begin{equation}
\Phi_{2,0}=\frac{1}{2}(\dot{y}^2+\omega_2^2y^2)
\end{equation}
and $\Phi_{i,n}$ polynomials of degree $n+2$. In the next section we describe the algorithm we use for the construction of the  third integral, that can be implemented with a computer algebra system. Here we use Maple.  For more details about the calculation of higher order terms of the third integral both in autonomous and non autonomous system see  \cite{Giorgilli1979,ContopMouts1966,Contopoulos1966a,Gustavson1966,Efthymiopoulos2005,tzemos2021integrals,tzemos2021order,tzemos2024formal}.

The series (3) is not convergent in general but 
it remains approximately constant in time when the perturbation $\epsilon$ is small. Convergent integrals were considered by several authors \cite{Cherry1926,Moser1956,Moser1958,Giorgilli2001},

In the most general case we have $\omega_1/\omega_2=$ irrational. 
But there are particular forms of the third integral 
with $\omega_1/\omega_2$  rational \cite{Contopoulos1963,ContopMouts1966,Contopoulos1966a}. A very special case 
is when $\omega_1=\omega_2=1$. This case has been studied 
in detail when  the perturbation is $H_1=-xy^2$ \cite{ContopMouts1966}.

However, no particular study has been made in the case 
of the H\'{e}non-Heiles system \cite{Henon1964} where the perturbation is 
$H_1=(xy^2-x^3/3)$ although special studies of the 
orbits in such a system have been made \cite{Nordholm1974,Waite1981,conte2005explicit,Takatsuka2022}. It is well 
known that when $\epsilon$ is small most orbits are 
regular but beyond a critical value of $\epsilon_{cr}$ most orbits are chaotic.

From the general theory it is known that chaos appears 
near every unstable periodic orbit \cite{contopoulos2002order,wiggins2003introduction}. In particular, the asymptotic 
curves from every unstable orbit intersect at an infinity 
of `homoclinic points'. But when the asymptotic curves of 
distant periodic orbits intersect at `heteroclinic 
points' then chaos becomes important. This is the 
phenomenon of `resonance overlap' \cite{Rosenbluth1966,Contopoulos1966c} and is 
responsible for the abrupt increase of chaos in 
generic non integrable systems. 

In the present paper we study the formal integrals 
and the appearance of chaos by resonance overlap in the H\'{e}non-Heiles system.

A remarkable phenomenon is that if we change the sign 
n the H\'{e}non-Heiles perturbation we find an integrable 
system with perturbation $H_1=(xy^2+x^3/3)$ \cite{chang1982analytic,fordy1991henon,lakshmanan1993painleve,lakshmanan2012nonlinear}. This
has been studied recently both from a classical and a Bohmian quantum perspective \cite{contopoulos2024}. (See also \cite{sengupta1996quantum,chattaraj2008quantum} for further work on the correspondence between the classical and the Bohmian quantum trajectories of the H\'{e}non-Heiles system.)

Here we study the transition from the 
integrable to the non integrable H\'{e}non-Heiles system 
by adding a perturbation $H_1=(xy^2+ax^3)$.

The structure of the paper is the following:
In section 2 we consider the integrals of motion of the Hamiltonian $H = \frac{1}{2}\left(\dot{x}^2 + \dot{y}^2 + x^2 + y^2\right) + \epsilon\left(xy^2 + \alpha x^3\right)$ that generalize the original H\'{e}non-Heiles Hamiltonian which has $\alpha = -1/3$. We give the invariant curves of the obtained surfaces of section as the values of $\epsilon$ and $\alpha$ change. We compare them with the exact surfaces of section found by numerical integration of many orbits. The exact surfaces of section have large chaotic regions when the perturbation $\epsilon$ approaches the escape perturbation for $\alpha<0$ \cite{zhao2007threshold,blesa2012escape}.
\par In section 3 we consider particular cases. Namely in section 3.1 we describe the integrable system with $\alpha = 1/3$, while in section 3.2 we provide details for the system with $\alpha = 0$. In section 3.3 we consider the curves with $\alpha > 0$, for $\alpha < 1/3$ and $\alpha > 1/3$. In section 3.4 we consider cases with  perturbations of the form $\epsilon\left(xy^2 + \alpha x^3 + \beta x^2 y + \gamma y^3\right)$. There we  find that if $\beta$ and/or $\gamma$ are different from zero  then we cannot construct a formal integral of motion because the terms which are linear in time $t$ (secular terms)  cannot be eliminated.
\par In section 4, we study the onset of chaos by the overlapping of many resonances in the original H\'{e}non-Heiles system. We give particular examples when the asymptotic curves of the unstable periodic orbits (UPOs) intersect at homoclinic and heteroclinic points.
\par Finally in section 5 we summarize our results and draw our conclusions.

\section{Approximate (formal) integrals of motion in the generalized H\'{e}non-Heiles systems}

From basic theory we  know that the integral of motion $\Phi$ given by \eqref{third}  must obey, for every order $n$ of $\epsilon$ the equation
\begin{equation}
    \frac{d\Phi_n}{dt}=[\Phi_n,H]=0.
\end{equation}
where
\begin{eqnarray}
    [\Phi_n,H]=\left(\frac{\partial \Phi_n}{\partial x}\frac{\partial H}{\partial \dot{x}}-\frac{\partial \Phi_n}{\partial \dot{x}}\frac{\partial H}{\partial x}\right)+\left(\frac{\partial \Phi_n}{\partial y}\frac{\partial H}{\partial \dot{y}}-\frac{\partial \Phi_n}{\partial \dot{y}}\frac{\partial H}{\partial y}\right)\label{eqint}
\end{eqnarray}
is the Poisson bracket.
Thus if we set the Hamiltonian \eqref{hams} in \eqref{eqint} then we can sort in ascending order the terms corresponding to the various powers of the perturbation $\epsilon$ and set them equal to $0$. Namely 
\begin{equation}\label{pde}
    \frac{\partial \Phi_{n+1}}{\partial t}+[\Phi_{n+1},H_0]-K_n=0,
\end{equation}
with
\begin{equation}\label{Kn]}
    K_n\equiv -\sum_{i=0}^n [\Phi_{n-i}, H_{i+1}]% = -[\Phi_n,H_1]-\dots-[\Phi_0,H_{n+1}], \quad n\geq 0.
\end{equation}
and $\Phi_0=\Phi_{1,0}+\Phi_{2,0}=H_0$.
By solving the PDE \eqref{pde} with the method of characteristics \cite{olver2014introduction} we find that 
\begin{equation}
    \Phi_{n+1}=\int K_n dt.\label{intK}
\end{equation}
For a Hamiltonian up to first order in $\epsilon$, this takes the form
\begin{equation}
    \Phi_{n+1} = \int [H_1, \Phi_n]\, dt
\end{equation}
Here we consider a Hamiltonian which is up to first order in $\epsilon$ and has the form 
\begin{equation}
H = H_o + \epsilon H_1 = \frac{1}{2}\left(\omega_x^2x^2 + \dot{x}^2 + \omega_y^2y^2 + \dot{y}^2\right) + \epsilon(xy^2+ax^3)\,.
\end{equation}
Thus $H_n=0,\,\, \forall n\geq 2$.
For non commensurable frequencies $\omega_x,\omega_y$ (non resonant case) the successive application of \eqref{intK} provides us with the various terms of the third integral.

However, in the present paper we work with $\omega_x=\omega_y=1$, i.e. we have a resonant case. Then a direct application of \eqref{intK} with $\Phi_0=H_0$ as the zero order integral will produce terms in the third integral that go to infinity \cite{contopoulos2002order}. In these cases if we set the value of the frequencies before performing the corresponding integration \eqref{intK} then we find that there exist terms containing time in linear form. These are the so called `secular terms' that need to be avoided for $\Phi$ to be an integral of motion. 

In order to proceed in the present  resonant case, the solutions of zero order in $\epsilon$ can be written as:
\begin{eqnarray}
\nonumber x=2\Phi_{1,0}\sin(t-t_0), \quad y=2\Phi_{2,0}\sin(t)\,,\\
\dot{x}=2\Phi_{1,0}\cos(t-t_0), \quad \dot{y}=2\Phi_{2,0}\cos(t)\,.\label{trigonometric}
\end{eqnarray}
Besides the zero order integrals $\Phi_{1,0}$ and $\Phi_{2,0}$
there are 3 more zero order integrals

\begin{align}
&C_0=(\dot{x}^2-x^2)(\dot{y}^2-y^2)+4\dot{x}\dot{y}xy=(2\Phi_{1,0})(2\Phi_{2,0})\cos(2t_0),\\
&Q_0=\dot{x}\dot{y}+xy=\sqrt{(2\Phi_{1,0})(2\Phi_{2,0})}\cos(t_0),\\
&P_0=\dot{x}y-\dot{y}x=\sqrt{(2\Phi_{1,0})(2\Phi_{2,0})}\sin(t_0).
\end{align}

In fact, there are 3 independent integrals: 
$\Phi_{1,0},\Phi_{2,0}$ and $C_0$. The first order terms in $\epsilon$ of these integrals are given by the formula 
\begin{eqnarray}
\Phi_{i,1}=\int\left(\frac{\partial \Phi_{i,0}}{\partial \dot{x}}\frac{\partial H_1}{\partial x}+\frac{\partial \Phi_{i,0}}{\partial \dot{y}}\frac{\partial H_1}{\partial y}\right)dt\,,
\end{eqnarray}
where 
\begin{eqnarray}
\frac{\partial H_1}{\partial x}=y^2+2ax, \quad \frac{\partial H_1}{\partial y}=2xy\,.
\end{eqnarray}
Thus 
\begin{eqnarray}
\Phi_{1,1}=\int \dot{x}(y^2+ax^2)dt, \quad \Phi_{2,1}=\int 2\dot{y}xydt.
\end{eqnarray}
In order to perform the integration we follow 3 steps:
\begin{enumerate}
\item We replace the variables $x,y,\dot{x},\dot{y}$
by their trigonometric forms \eqref{trigonometric} and write them as
sums of simple trigonometric terms of the form $\mathrel{\substack{\text{cos} \\ \text{sin}}}(m(t-t_0)-nt)
$. 
\item We integrate by simply replacing $\cos$ and $\sin$ with $\mathrel{\substack{\text{cos} \\ \text{sin}}}\left(m(t-t_0)-nt\right)/(m-n)$. In general
 $m\neq n$. But if $m=n$ then we have a secular term 
 $\mathrel{\substack{\text{cos} \\ \text{sin}}}(m t_0)\, t$\, which increases indefinitely in time.
\item The resulting trigonometric expression is expressed 
in terms of $x,y,\dot{x},\dot{y}$ by using again \eqref{trigonometric} for the inverse transformation.
\end{enumerate}

Further details are given in the paper \cite{ContopMouts1966}. In the present case we have a problem when we encounter a secular term of the form $\sin(mt_0)\,t$. Then we try to find a combination of the independent integrals that does not produce secular terms. 

In particular, if we calculate the first order terms in $\epsilon$ $\Phi_{1,1},\Phi_{2,1}$, (which are of third degree in the variables $x,y,\dot{x},\dot{y}$)  we find that they are free of secular terms. However, if we calculate  the second order terms in $\epsilon$ then we  find secular terms of  degree 4 in the variables containing the factor $\sin(2t_0)t$. On the other hand, the first degree in $\epsilon$ terms of $C_1$ are found to be of degree 5 in the variables with no secular terms, but the next term, $C_2$ has secular terms of degree 6 containing the factor $sin(2t_0)\, t$.

However, if we start with zero order integrals $(2\Phi_{1,0})^2$,$(2\Phi_{1,0})(2\Phi_{2,0})$, $(2\Phi_{2,0})^2$ we find secular terms of degree 6 which are of the same form as the corresponding secular terms of $C_2$. Thus we expect that a combination of the zero order terms 
\begin{equation}
\phi_0=C_0+c_1(2\Phi_{1,0})^2+c_2(2\Phi_{1,0})(2\Phi_{2,0}) +c_3(2\Phi_{2,0})^2\label{comb}
\end{equation}
will be able to avoid the secular terms of order $\epsilon^2$. This in fact what we manage to do.

%In fact we can eliminate in the same way the higher order in $\epsilon$ by using appropriate powers of $\Phi_{1,0}$ and $\Phi_{2,0}$ [ref].

In the present case the calculations give the terms of order $\epsilon$,
\begin{align}
&\Phi_{1,1}=\frac{1}{3}\left(xy^2+2x\dot{y}^2-2y\dot{x}\dot{y}+3ax^3\right)\,,\\&
\Phi_{2,1}=H_1-\Phi_{1,1}
\end{align}
and
\begin{equation}
    \begin{aligned}
    C_1=&-\frac{1}{3}(4+6a)x^3y^2+(4-6a)x^3\dot{y}^2+12(1-a)xy^2\dot{x}^2\\
    &+2xy^4-6xy^2\dot{y}^2-12(1-2a)xy^2\dot{x}^2\\
    &-4{x}\dot{y}^4+12(1-2a)x\dot{x}^2\dot{y}^2-8y^3\dot{x}\dot{y}+4y\dot{x}\dot{y}^4-12(1-2a)y\dot{x}^3\dot{y}.
    \end{aligned}
\end{equation}
Thus the total first order term 

\begin{equation}
\phi_1=C_1+4\Big[c_1(2\Phi_{1,0})\Phi_{1,1}+c_2\left[(2\Phi_{1,0})\Phi_{2,1}+(2\Phi_{2,0})\Phi_{1,1}\right]
\\+c_3(2\Phi_{2,0})\Phi_{2,1}\Big]\label{C1}
\end{equation}
These terms are free of secular terms. However, in the second approximation (of order $\epsilon^2$) we find the following secular terms. 

%In this approximation $C_2$ is of degree 6 in the variables $x,y,\dot{x}\dot{y}$, while $\Phi_{02}$ and $\Phi_{1,2}$ are of degree 4. In order to reach degree 6 we have to take terms of the form $c_1(2\Phi_{1,0})^2+c_2(2\Phi_{1,0})(2\Phi_{2,0})+c_3(2\Phi_{2,0}^2)$. In order to eliminate the secular terms of $C_2$. In fact the secular terms of an integral $\phi=\phi_0+\epsilon\Phi_1+\epsilon^2\phi_2$ where $\phi_0=C_0+c_1(2\Phi_{1,0})^2+c_2(2\Phi_{1,0})(2\Phi_{2,0})+c_3(2\Phi_{2,0}^2)$ are
\begin{equation}
\begin{aligned}
\epsilon^2\Phi_{2,\text{sec}}=&\frac{\epsilon^2}{6}\Big([18\alpha-1-3(2-\alpha)c_2+6(2-\alpha)c_3]\bar{A}\\
&+[45\alpha^2-18\alpha-4-6(2-\alpha)c_1+3(2-\alpha)c_2]B\Big)t
\end{aligned}
\end{equation}
where
\begin{equation}
\bar{A}=(2\Phi_{1,0})^2(2\Phi_{2,0})\sin(2t_0), \quad \bar{B}=(2\Phi_{1,0})(2\Phi_{2,0})^2\sin(2t_0)\,.
\end{equation}
These terms are zero if
\begin{equation}
\begin{aligned}
&c_1=\frac{45\alpha^2-18\alpha-4}{6(2-\alpha)}\,,\\
&c_2=0\\
&c_3=\frac{1-18\alpha}{6(1-\alpha)}\,.
\end{aligned}
\end{equation}

Higher order terms can be calculated in the same way. In particular the term $\epsilon^3\Phi_3$ does not contain secular terms while $\epsilon^4\Phi_4$ contains secular terms which can also be avoided as shown in \cite{ContopMouts1966}. In fact, in that paper we calculated the third integral up to order 8. The higher order terms are very lengthy and cannot be given here. In present paper we truncate  the third integral at second order in $\epsilon$.

The form of the integral
\begin{equation}
\phi=\phi_0+\phi_1\epsilon+\phi_2\epsilon^2
\end{equation}
without secular terms in $\phi_2$ is used in order to find the invariant curves on a surface of section at $y=0$. In order to do that we must replace $\dot{y}^2$ with its value from the Hamiltonian equation. Namely for $y=0$
\begin{equation}\label{ydot2}
\dot{y}^2=2-(\dot{x}^2+x^2)-2\epsilon \alpha x^3\,.
\end{equation}
Then we find an equation that gives the invariant curves from any initial point $(x_0,\dot{x}_0$) for $y_0=0$ and $E=1$, in the form 
\begin{equation}
f(x,\dot{x})=f(x_0,\dot{x}_0)\,.
\end{equation}
We must also have $\dot{y}^2\geq 0$. Thus the invariant curves must have $\dot{x}^2$ smaller than
\begin{equation}
\dot{x}^2=2 - x^2 - 2\epsilon\alpha x^3\label{xdot2}
\end{equation}
This equation has three roots $x_1$, $x_2$, $x_3$ for $\dot{x}^2 = 0$, which are real if $\epsilon$ is small. If $\alpha > 0$ we have $x_1 > 0 > x_2 > x_3$ and $\dot{x}^2$ is positive if  $x_2<x<x_1$,  as well, as if $x<x_3$ . If $\alpha < 0$ then $x_1 < 0 < x_2 < x_3$ and $\dot{x}^2$ is positive for $x_1<x<x_2$ and for $x > x_3$. But at a critical value of $\epsilon = \epsilon_\text{esc}$ (escape perturbation) the roots $x_2$ and $x_3$ coincide. For $\epsilon>\epsilon_{esc}$ there is only one real root, $x_1$. The escape perturbation is found if the equation \eqref{xdot2} for $\dot{x}^2 = 0$ has a double root something that happens when its derivative is zero, i.e. when $x = -\frac{1}{3\epsilon\alpha}$. Then the equation $2 - x^2 - 2\epsilon\alpha x^2 = 0$ gives
\begin{equation}
\epsilon_{esc}=\frac{1}{3\sqrt{6}|\alpha|}.
\end{equation}
E.g. for $\alpha=\pm \frac{1}{3}$, $\epsilon_{esc}=0.408$.
This curve is close to the  circle $\dot{x}^2+x^2=2$, which is  larger along the $\dot{x}=0$ axis to the right if $\alpha<0$ or to the left if $\alpha>0$. If $\epsilon$ is fixed then there is a corresponding escape value of $\alpha$  equal to $\alpha_{esc}=\frac{1}{3\sqrt{6}\epsilon}$.

Another restriction refers to escapes along the y-direction. In fact, the orbits on the $(x,y)$ plane must be inside the curve of zero velocity (CZV) on the $x-y$ plane, where $\dot{x}=\dot{y}=0$. This is given by the equation
\begin{equation}
y^2=\frac{2-x^2-2\epsilon \alpha x^3}{1+2\epsilon \alpha x}\,.
\end{equation}
The above equation gives $y^2\to\infty$ if $x=-\frac{1}{2\epsilon}$ with $2-x^2-2\epsilon \alpha x^3>0$
By setting this value of $x$ in the equation $2-x^2-2\epsilon \alpha x^3=0$ we find a new escape perturbation 
\begin{equation}
\epsilon_{esc}'=\sqrt{\frac{1-\alpha}{8}}\,.
\end{equation}
E.g. for $\alpha=0$ we have $\epsilon_{esc}'=0.354$ and for $\alpha=-1/3$, $\epsilon_{esc}'=0.408$ i.e. the same escape value as $\epsilon_{esc}$. For $\alpha=1/3$ we find $\epsilon_{esc}'=0.289$. If $\epsilon$ is fixed then the corresponding escape $\alpha$ is $\alpha_{esc}=1-8\epsilon^2$.
\par The value of $\epsilon_\text{esc}^\prime$ is smaller that $\epsilon_\text{esc}$ for every $\alpha > -1/3$. In these cases escapes towards $y = \pm \infty$ appear as $\epsilon$ increases from $\epsilon=0$ with $0<\epsilon<\epsilon_\text{esc}$.

\begin{figure}[H]
\centering
\includegraphics[scale=0.3]{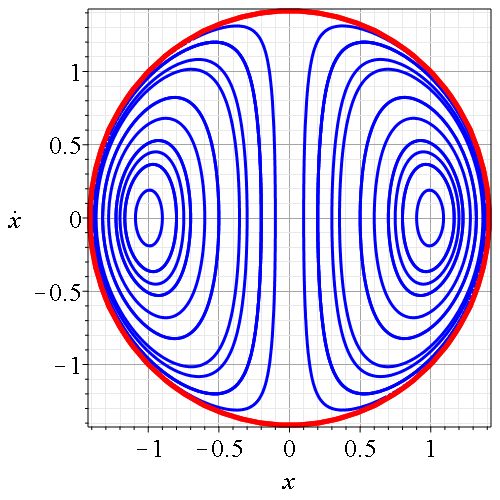}
\includegraphics[scale=0.3]{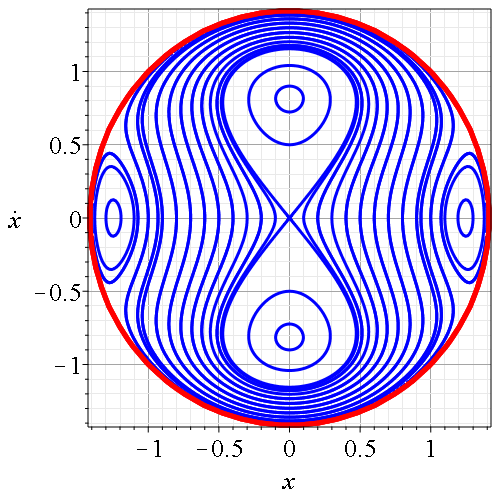}
\caption{Surfaces of section of the theoretical integrals in the unperturbed cases ($\epsilon = 0$) for (a) $\alpha = -1/3$ (non-integrable H\'{e}non-Heiles Hamiltonian) and (b) $\alpha = 0$. The boundaries (red curves) correspond to the curve $\dot{y}^2=0$.}\label{fig:eps0}
\end{figure}

In the following figures we show the invariant curves on the surface of section $y=0$ for various values of  $\alpha$ and $\epsilon$.
In \figref{eps0} we give the limiting surfaces of section for $\epsilon=0$ (a) $\alpha=0$ and (b) $\alpha=-1/3$ (non integrable H\'{e}non-Heiles). In \figref{eps0}a there are only two stability around stable periodic orbits (on the $x$-axis, symmetric with respect to the line $y=0$. In \figref{eps0}b there are two asymmetric islands on the x-axis but there are also two islands above and below the origin separated by a UPO at the origin $(0,0)$.

\begin{figure}[H]
\centering
\includegraphics[scale=0.3]{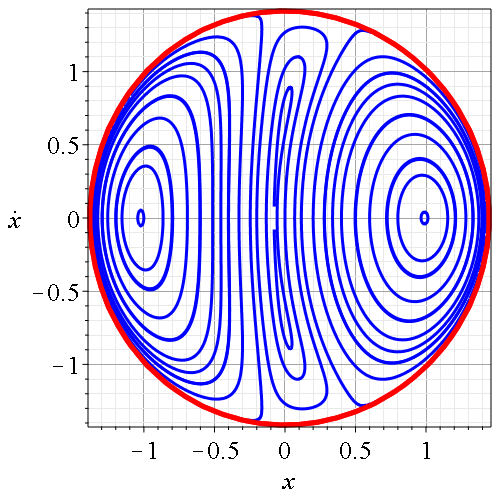}
\includegraphics[scale=0.45]{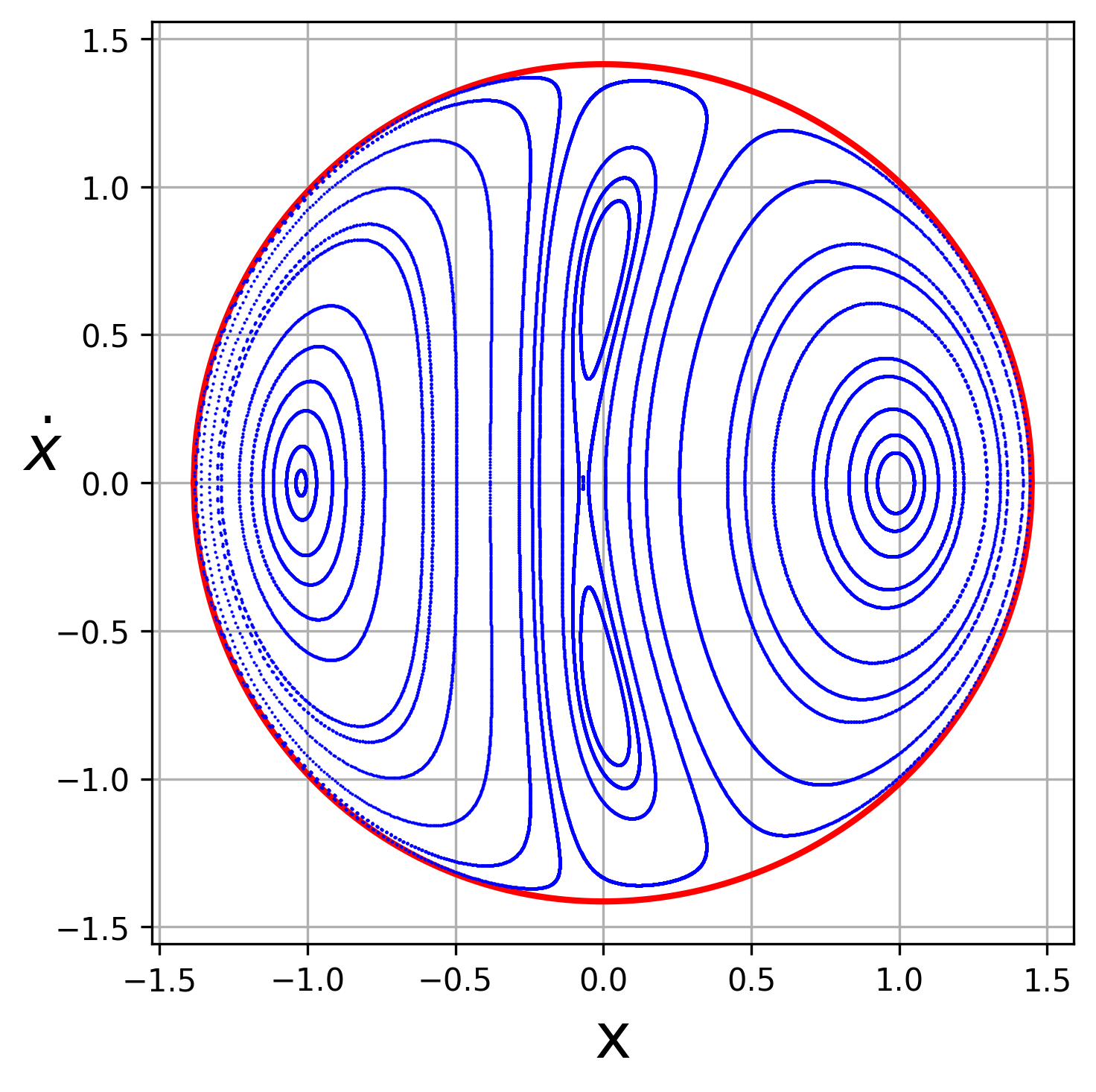}
\caption{Theoretical (a) and exact (numerical) (b) surfaces of section in the case $\epsilon = 0.05$, $\alpha = -0.32$.}\label{fig:005minus032eps0}
\end{figure}

We continue  with the cases where $\epsilon\neq0$. There we give the invariant curves  on the surface of section $y=0$ defined by the integral truncated at order $\epsilon^2$ and the exact numerical invariant curves found by integrating orbits for many initial conditions. The boundaries of the figures (red curves) represent the limiting curves of Eq.~\eqref{ydot2} for $\dot{y}^2=0$. If we truncate the integrals at order $\epsilon$ then we generally find a worse agreement with the exact curves, except in the integrable case with $\alpha = 1/3$ (see section 3.1). 
In particular, in \figref{005minus032eps0} we give the invariant curves for a small value of $\epsilon$ ($\epsilon = \mathbf{0.05}$) and $\alpha=-0.32$ (theoretical and exact). There are two SPOs on the $x$-axis as in \figref{eps0}, but there are also two SPOs above and below the origin, separated by a UPO on the left close to the origin. The theoretical and the exact invariant curves are close to each other.

\begin{figure}[H]
\centering
\includegraphics[scale=0.3]{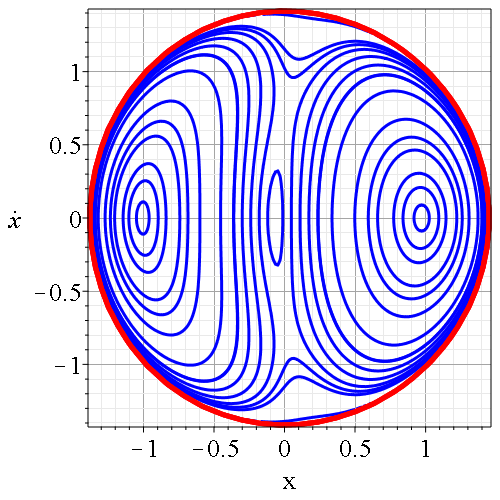}
\includegraphics[scale=0.45]{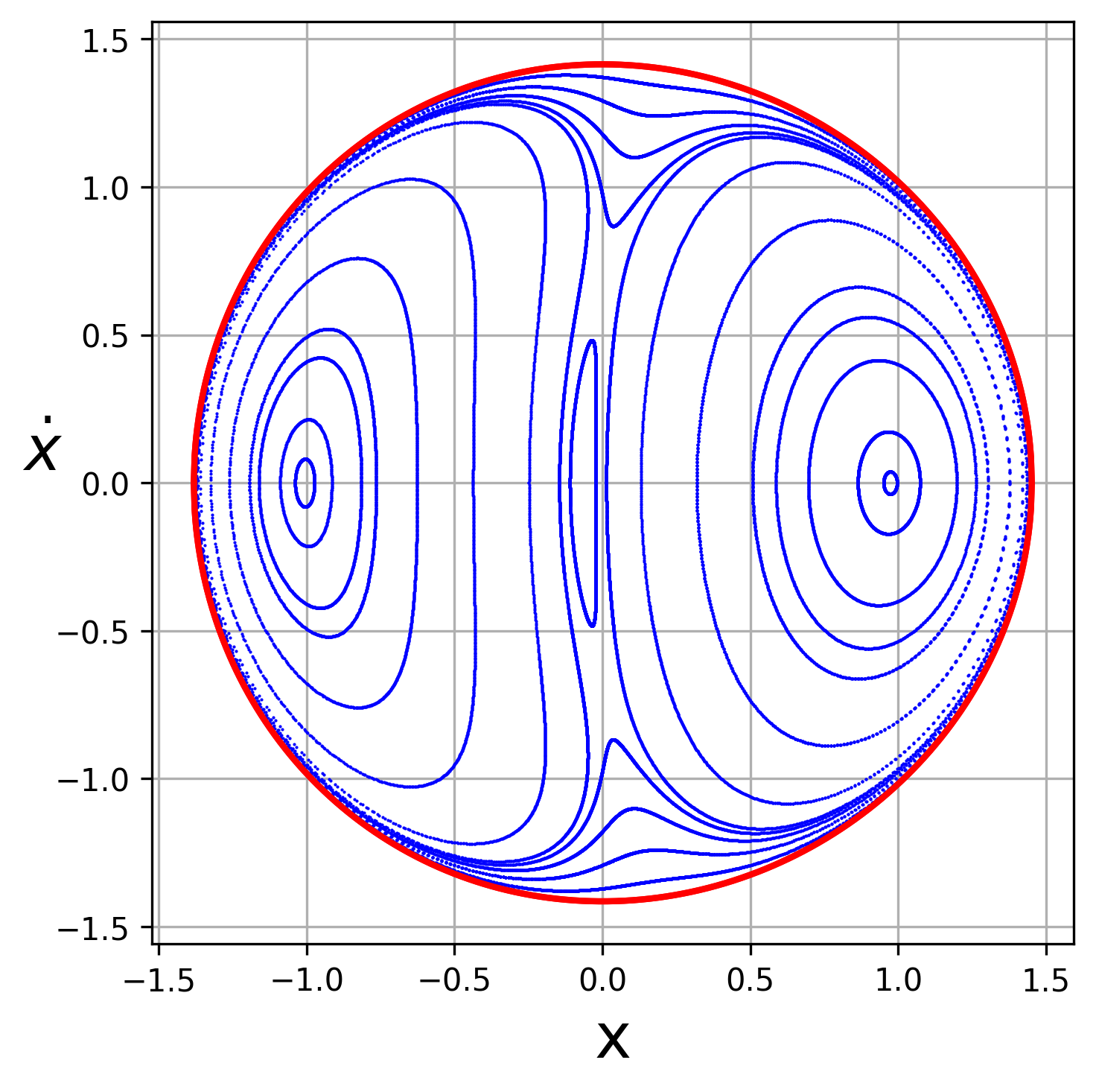}
\caption{Theoretical (a) and exact (b) surfaces of section of the case $\epsilon = 0.05$, $\alpha = -0.35$.}\label{fig:005minus035}
\end{figure}

As $\alpha$ becomes more negative ($\alpha=-0.35$) the theoretical and the numerical surfaces of section are again similar (\figref{005minus035}a,b), but they have a stable periodic orbit a little to the left of the origin and two UPOs above and below it.

The transition from unstable to SPOs for $\epsilon=0.05$ takes place at an $\alpha$ little larger than $\alpha=-1/3$ for the second order approximation of the third integral, and at a value of $\alpha$ a little smaller than $\alpha = -1/3$ in the exact case. Thus the truncated theoretical integral does not represent well the numerical surface of section at exactly  $\alpha = -1/3$.

\begin{figure}[H]
\centering
\includegraphics[scale=0.3]{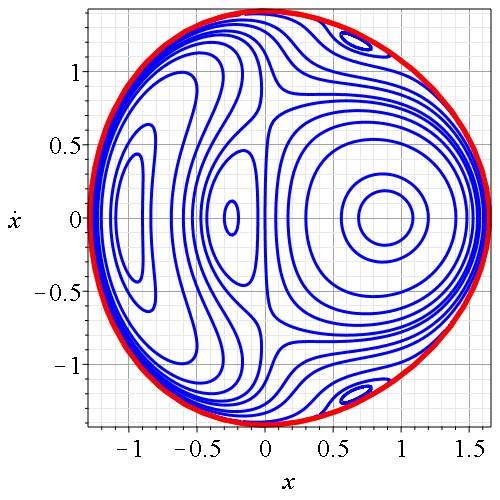}
\includegraphics[scale=0.45]{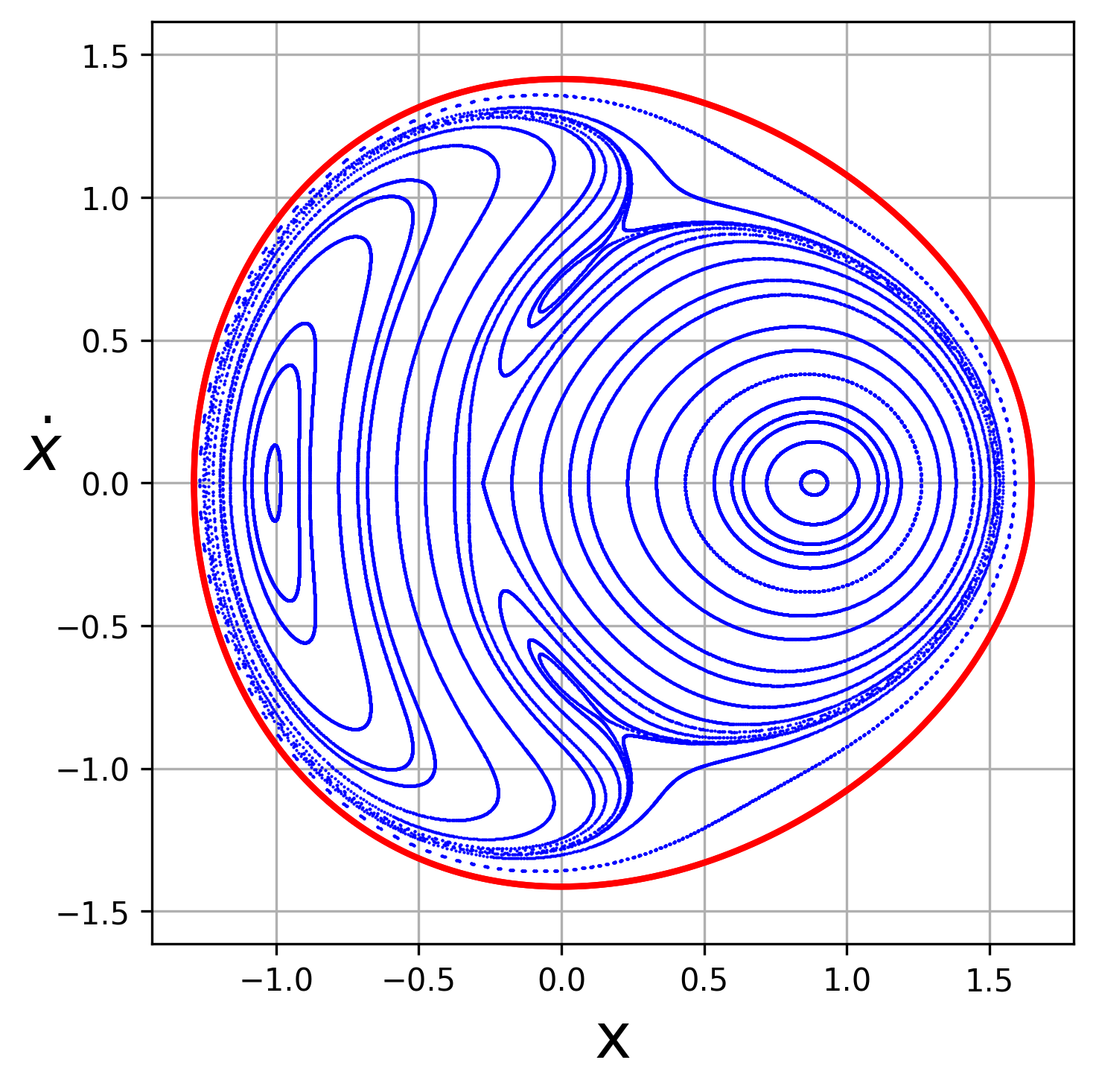}
\includegraphics[scale=0.45]{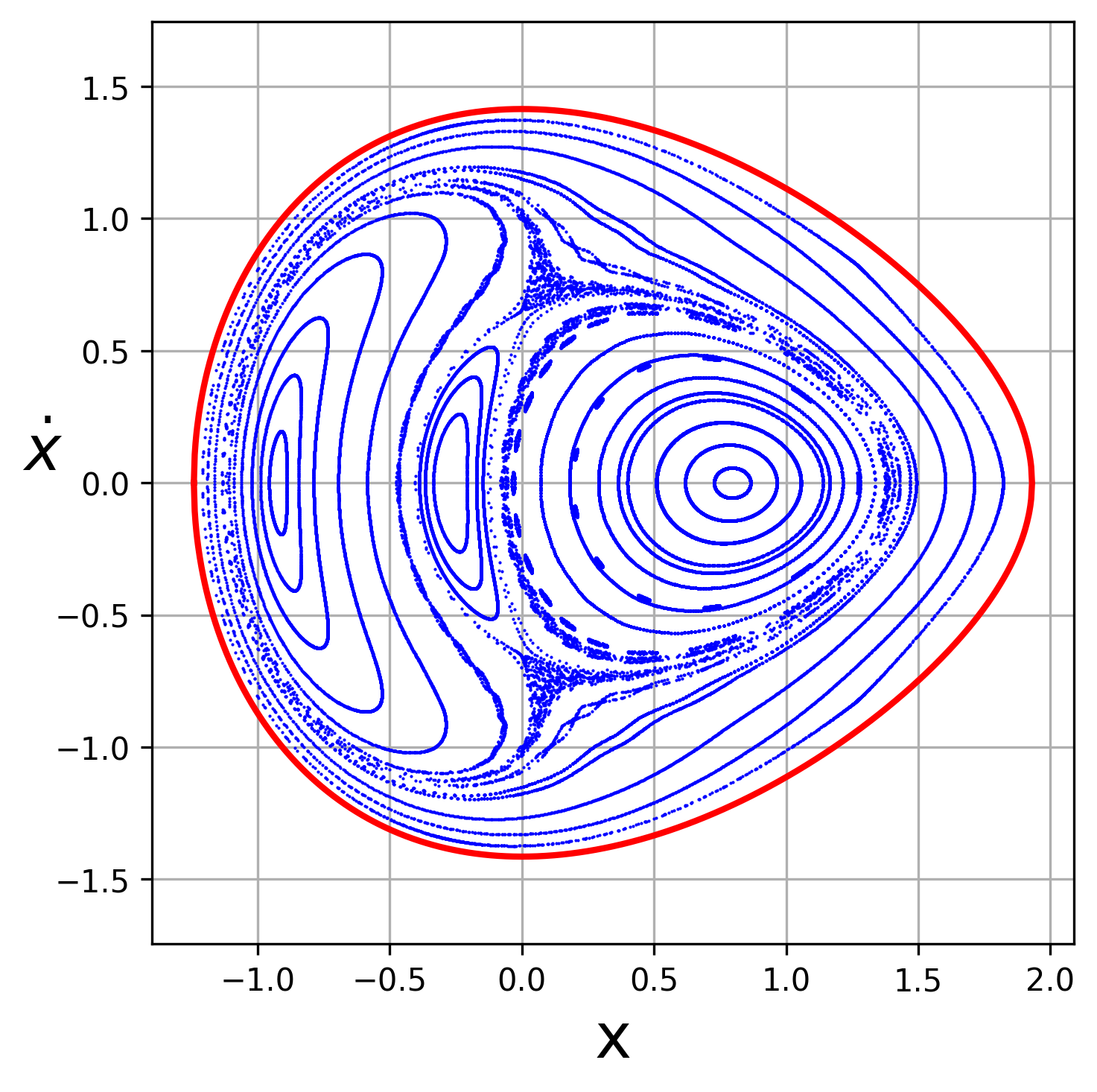}
\includegraphics[scale=0.45]{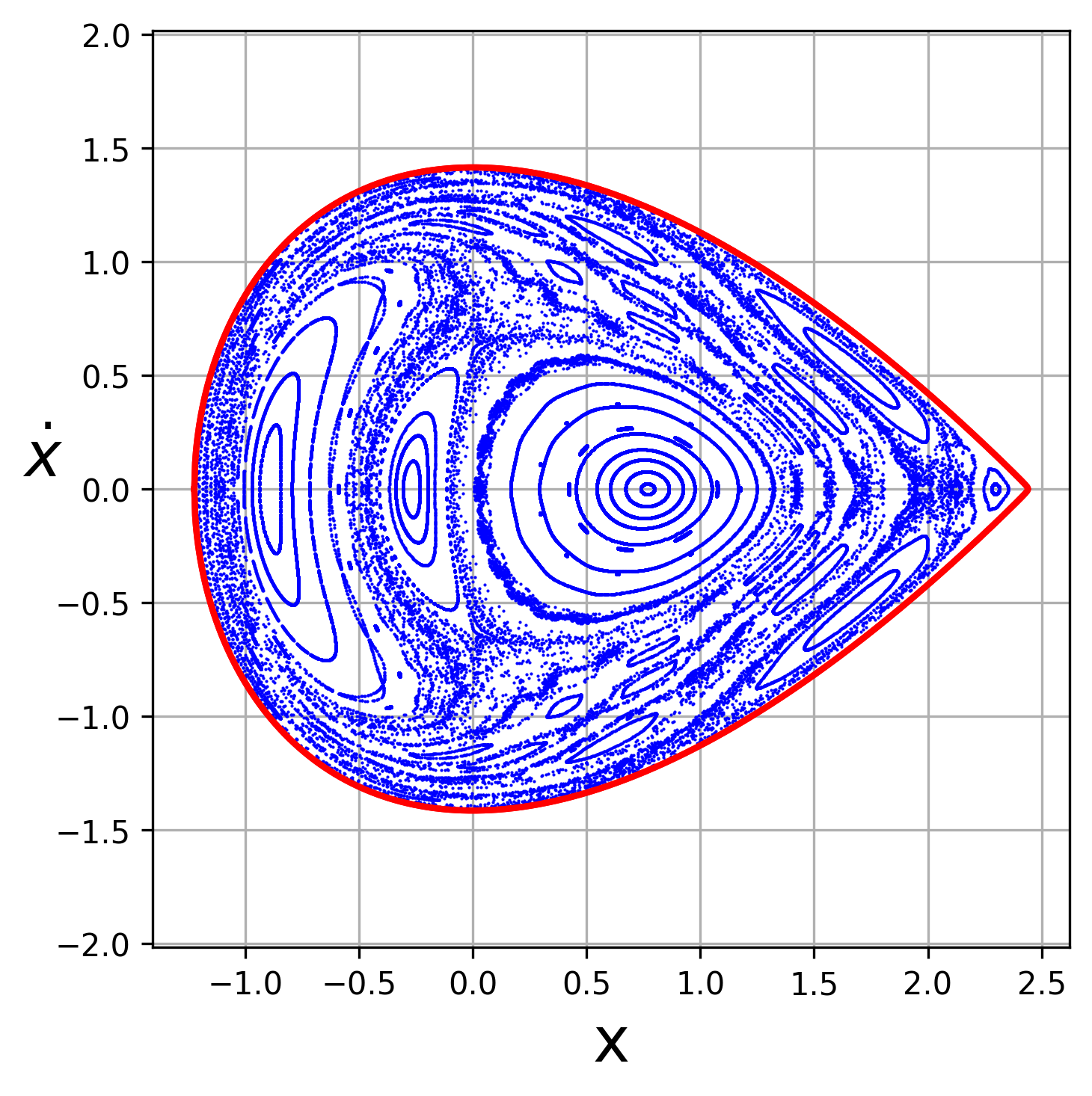}
\caption{Theoretical (a) and exact (b, c, d) surfaces of section of the cases with $\epsilon = 0.2$ and $\alpha = $ (a) $-0.4$, (b) $-0.4$, (c) $-0.6$, (d) $-0.6804$.}\label{fig:eps02}
\end{figure}

For larger $\epsilon$ the deviations between the theoretical and the exact numerical results become larger. For $\epsilon=0.2$ and $\alpha=-0.4$ the theoretical surface of section (\figref{eps02}a) contains a large island around a central stable periodic orbit on the left of the origin, while in the exact surface of section (\figref{eps02}b) the central periodic orbit is unstable. However for smaller $\alpha$ (e.g. for $\alpha=-0.6$) (\figref{eps02}c), the exact periodic orbit is stable. For even smaller $\alpha$ ($\alpha=0.6804$ \figref{eps02}d) this orbit remains stable theoretically, while in the exact case the largest part of the surface of section contains chaos. The last case is very close to the escape perturbation which for $\epsilon=0.2$ is $\alpha_{esc}=0.68041$.

\begin{figure}[H]
\centering
\includegraphics[scale=0.6]{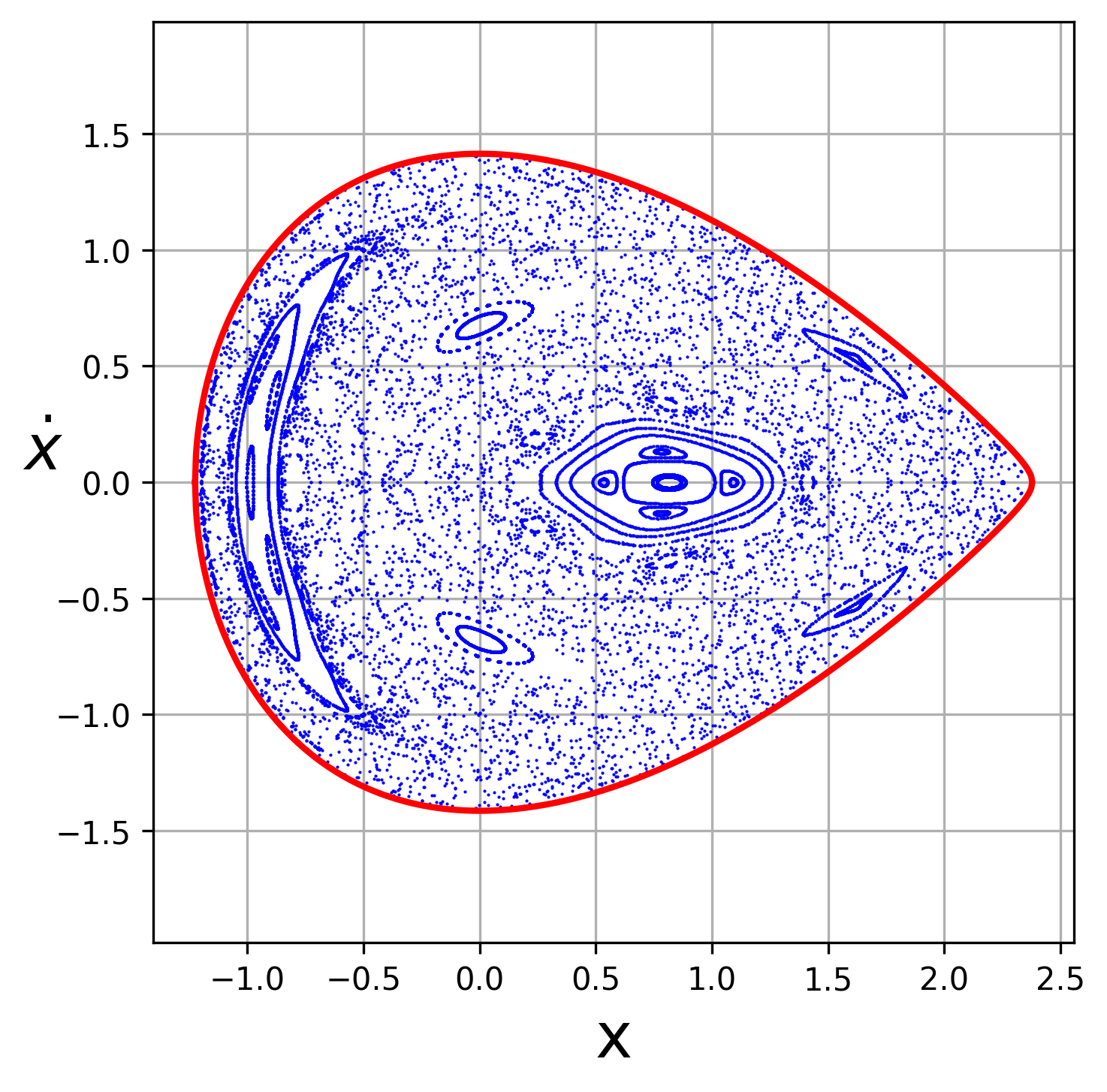}
\includegraphics[scale=0.6]{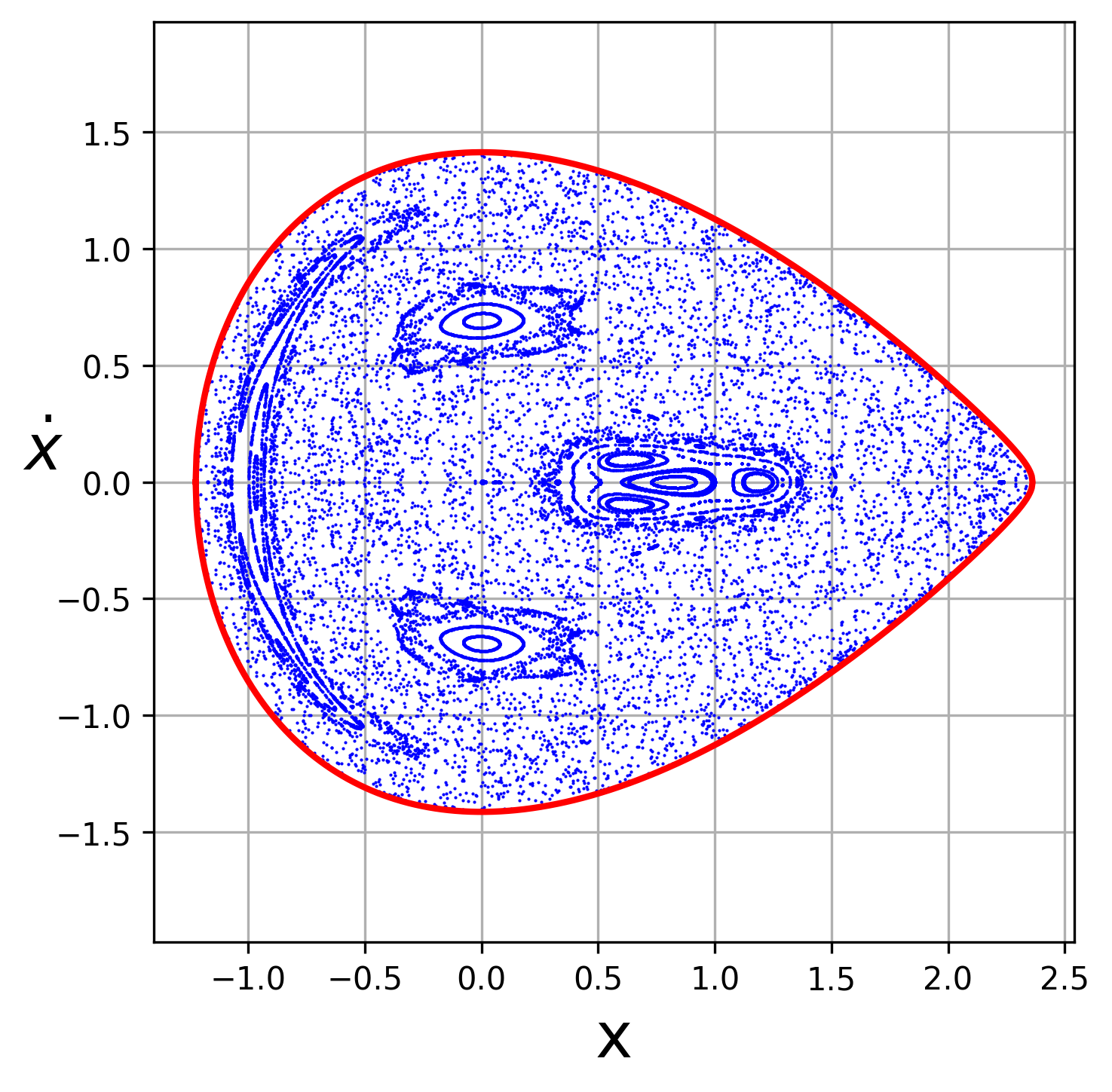}
\caption{Exact surfaces of section for (a) $\epsilon = 0.3$, $\alpha = -0.453$ and (b) $\epsilon = 0.35$, $\alpha = 0.388$.}\label{fig:chaotic}
\end{figure}

On the other hand, for $\epsilon=0.3$ and $\epsilon=0.35$ the exact surface of section does not contain an island around a stable periodic orbit near the center for all the values of $\alpha$ from $\alpha=-1/3$ up to the escape perturbation, which is $\alpha=0.4356$ for $\epsilon=0.3$ and $\alpha=0.3881$ for $\epsilon=0.35$. Instead, in the exact case we always have an unstable central periodic orbit for $x < 0$ and $\dot{x} = 0$ and a large degree of chaos around it (see \figref{chaotic}a for $\epsilon=0.3$ and $\alpha=-0.453$, and \figref{chaotic}b for $\epsilon=0.35$ and $\alpha=-0.388$). The corresponding theoretical surfaces of section have a large island near the center as in the case of \figref{eps02}a.

The most important deviation is due to the appearance of chaos in the exact surface of section which is most important around the UPO on the left of the origin. The onset of chaos will be discussed in Section 4.

\section{Particular cases of integrals}
\subsection{Integrable H\'{e}non-Heiles system}

A particular case is  the integrable H\'{e}non-Heiles system with $\alpha=\frac{1}{3}$.
Its Hamiltonian is 
\begin{equation}
H=\frac{1}{2}(\dot{x}^2+\dot{y}^2+x^2+y^2)+\epsilon \left(xy^2+\frac{x^3}{3}\right)=E.\label{ham}
\end{equation}
We work again with $E=1$. 

\begin{figure}[H]
\centering
\includegraphics[scale=0.45]{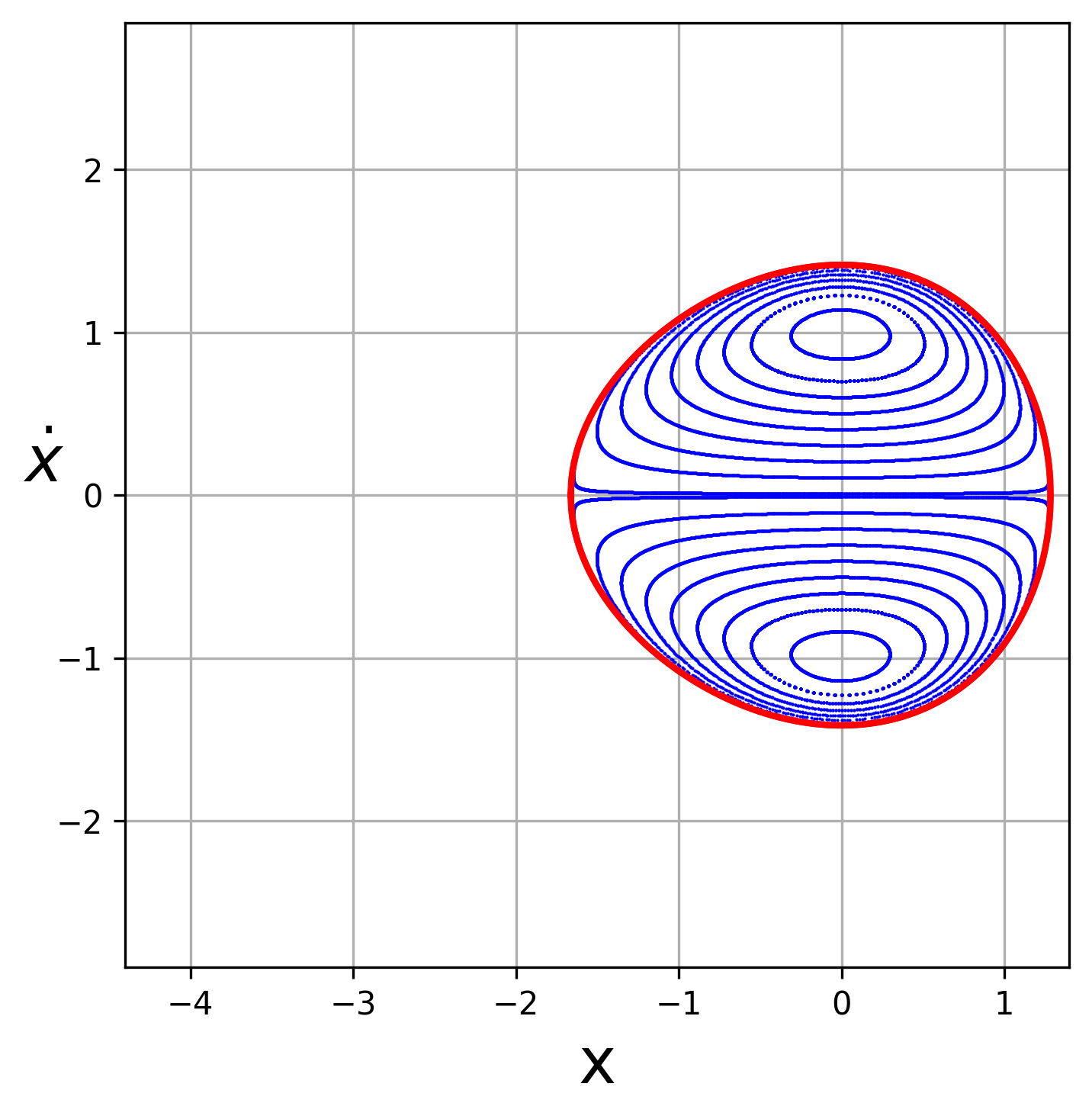}
\includegraphics[scale=0.45]{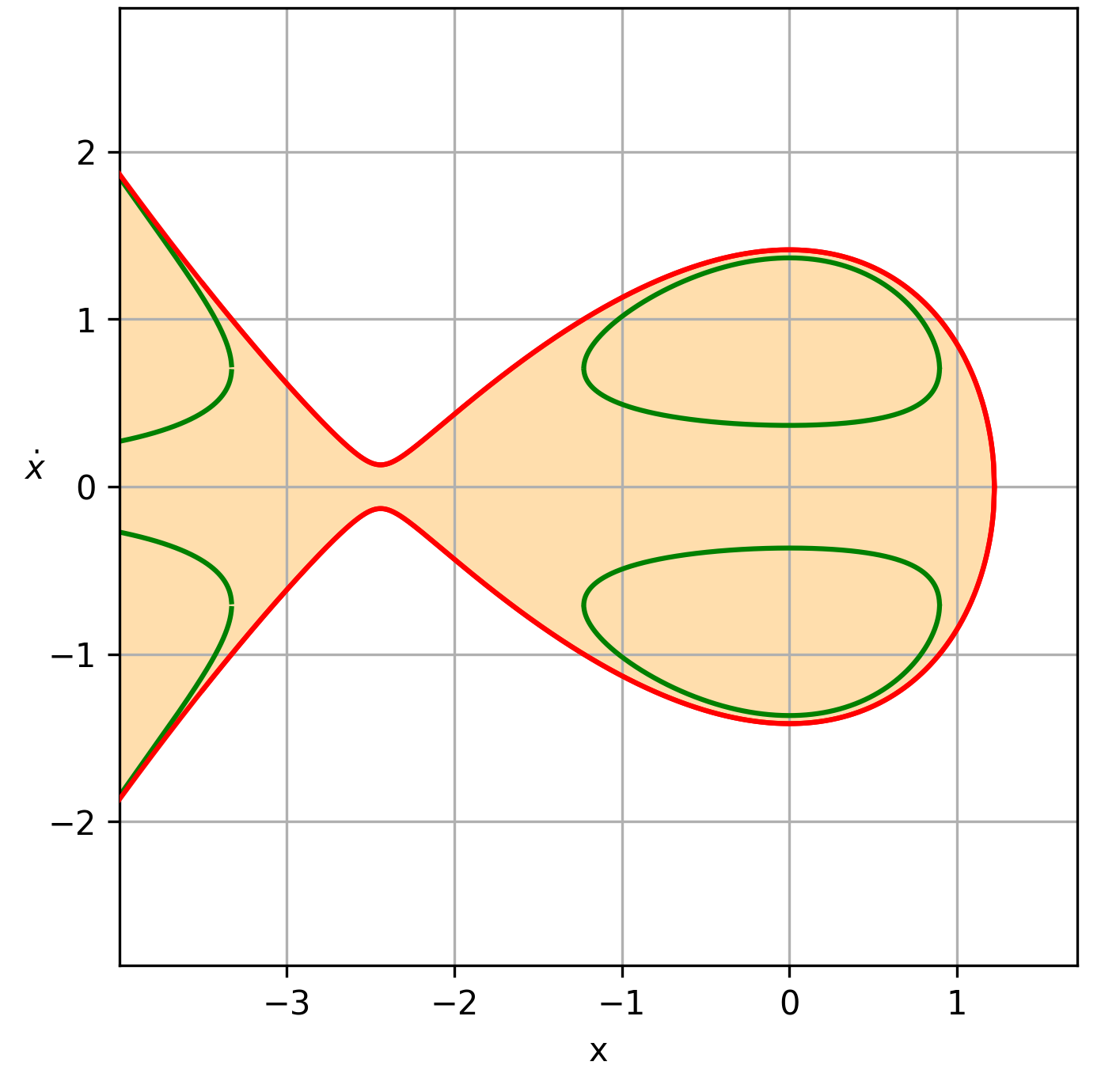}
\includegraphics[scale=0.45]{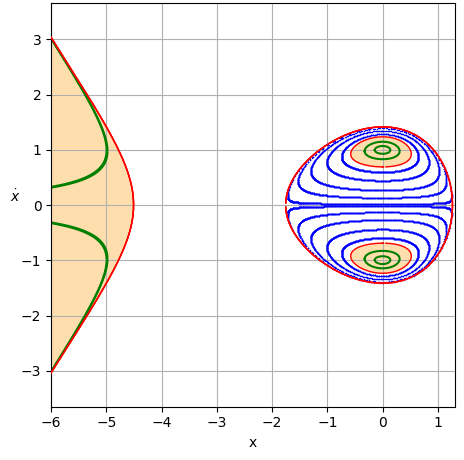}
\includegraphics[scale=0.3]{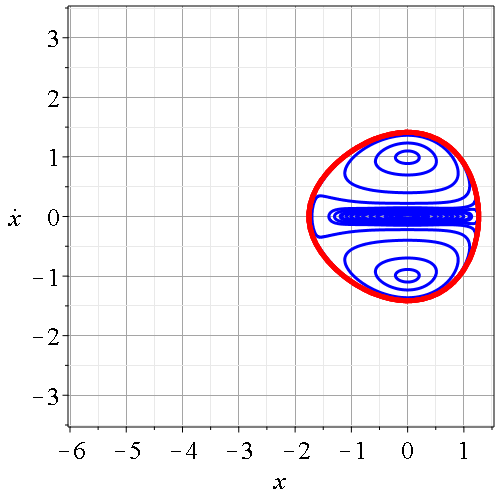}
\caption{Surfaces of section ($y=0$) in the integrable case $\alpha = 1/3$. (a) Case $\epsilon = 0.25$, (b) Case $\epsilon = 0.41$, (c) Case $\epsilon = 0.3$, (d) Theoretical case $\epsilon =  0.3$ when we use the integral $\Phi = \Phi_0 + \epsilon \Phi_1 + \epsilon^2 \Phi_2$. In the yellow regions (escaping orbits) we give some invariant curves for fixed values of $K$ corresponding to  the initial conditions of different escaping orbits (green). In (a) the left region where $\dot{x}^2>0$ is far away, outside \figref{figlabel}a for $x<x_3=-5.620$. In (c) this region is closer to the oval with $x<x_3=-3.608$. In (b) this region has joined the main oval, because $\epsilon=0.41$ is larger than the escape perturbation 
 $\epsilon_{esc}=0.408$.}\label{fig:figlabel}
\end{figure}

In this case we can use the zero order integral \cite{lakshmanan1993painleve,lakshmanan2012nonlinear}
\begin{align}
\Phi_{1,0}=Q_0=\dot{x}\dot{y}+xy.
\end{align}
Then the first order term in $\epsilon$ is 
\begin{equation}
Q_1 \equiv \epsilon\int \Big(\dot{y}(y^2+x^2)+\dot{x}(2xy)\Big)dt=\epsilon\left(yx^2+\frac{y^3}{3}\right).
\end{equation}
and does not contain any momentum ($\dot{x}$ or $\dot{y}$). Therefore the next term $Q_2$ is zero. In fact the integral is exact:
\begin{equation}
Q=\dot{x}\dot{y}+xy+\epsilon\left(yx^2+\frac{y^3}{3}\right)=K.
\end{equation}
and it is valid for any value of $\epsilon$.  The surface of section at $y=0$ gives
\begin{equation}
\dot{x}\dot{y}=K
\end{equation}
or $Q_0^2=\dot{x}^2\dot{y}^2=K^2$, where $\dot{y}$ is found from the integral \eqref{ham} by
\begin{align}
\dot{y}^2=2-\dot{x}^2-x^2-2\epsilon x^3/3.
\end{align}
If $\dot{y}^2=0$ then we have a limiting curve $\dot{x}^2=2-x^2-\frac{2\epsilon x^3}{3}$, which is  a circle for $\epsilon=0$ and it is an oval elongated to the left for $\epsilon > 0$. The corresponding invariant curves are
\begin{equation}
Q_0^2=-\dot{x}^4+\left(2-x^2-\frac{2\epsilon x^3}{3}\right)\dot{x}^2+K^2=0
\end{equation}
and give the solutions (\figref{figlabel}a,c)
\begin{equation}
\dot{x}^2=\frac{1}{2}\left[2-x^2-\frac{2\epsilon x^3}{3}\pm \sqrt{\left(2-x^2-\frac{2}{3}\epsilon x^3\right)^2-4K^2}\right],
\end{equation}
which are real if
\begin{equation}
\left(2-2x^2- \frac{2\epsilon x^3}{3}\right)^2\geq 4K^2.
\end{equation}

For $x=0$ we must have  $K^2\leq 1$ and for $K^2=1$ we find invariant points with $\dot{x}=\pm 1$ and the axis $\dot{x}^2=0$. Then for $K=0$ we have $\dot{x}=\pm \sqrt{2}$ and $0$.

The invariant curves for $\epsilon<\epsilon_{esc}$ are closed around the points ($x=0,\dot{x}=\pm 1$) if $F=2-x^2-2\epsilon x^3/3\geq 0$. Then for a given $x$ we have a maximum $\dot{x}^2=F$ for $K=0$. If we try to find orbits with $\dot{x}^2>F$ we find that $\dot{y}^2<0$. Therefore such orbits do not exist.

The limiting values of $x$ are the roots of the equation 
\begin{equation}
F \equiv 2-x^2-2\epsilon x^3/3=0
\end{equation}
and for $\epsilon<\epsilon_{esc}$ it has 3 real roots $x_1>0>x_2>x_3$. For example if $\epsilon = 0.25$ (\figref{figlabel}a) we have $x_1 \approx 1.284$, $x_2 \approx -1.664$, $x_3 \approx -5.620$, and if $\epsilon = 0.3$ (\figref{figlabel}c), we have $x_1 \approx 1.264$, $x_2 \approx -1.756$, $x_3 \approx -4.508$. The quantity $F$ is positive between $x_1$ and $x_2$, and positive again for $x < x_3$. In the interval between $x_2$ and $x_3$ $F$ is negative and the invariant curves do not reach the $\dot{x}=0$ axis.
\par In the region on the left of $x_3$ we have invariant curves for $y=0$ that extend to $\infty$ (\figref{figlabel}c).

 When $\epsilon$ reaches the escape value $\epsilon_\text{esc} = 0.488$ the roots $x_2$ and $x_3$ coincide and for larger $\epsilon$ they become complex. Then the regions where $F$ is positive join (\figref{figlabel}b) and the orbits escape to $x\rightarrow -\infty$.

There are also escaping orbits along the y-direction (in fact along the directions ($x \rightarrow -\infty$, $y \rightarrow \pm \infty$)) when $\epsilon > \epsilon_\text{esc}^\prime = \frac{1}{2\sqrt{3}} = 0.289$, as derived from Eq. 42. 

\begin{figure}[H]
\centering
\includegraphics[scale=0.33]{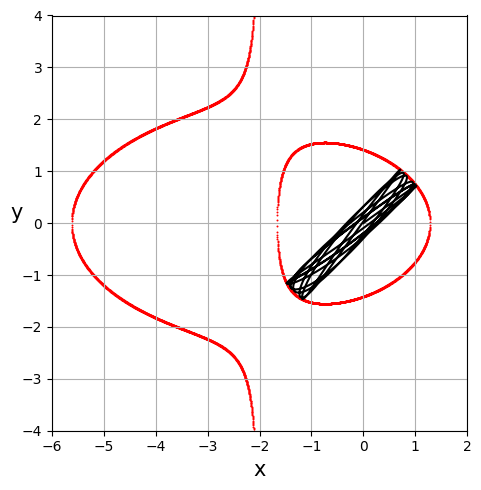}[a]
\includegraphics[scale=0.33]{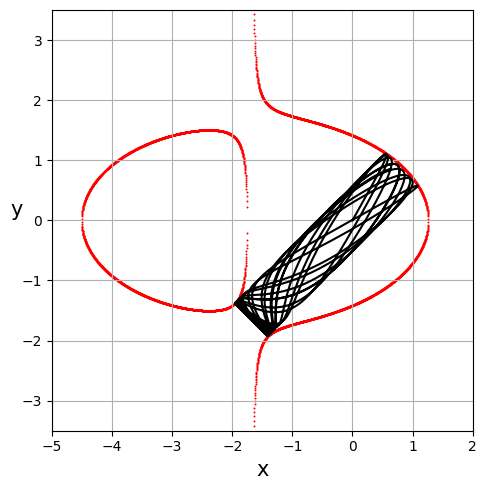}[b]
\includegraphics[scale=0.33]{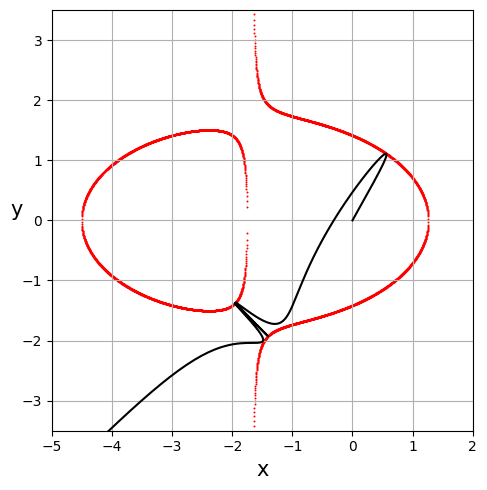}[c]
\caption{(a) A Lissajous orbit inside the closed part of the zero velocity curve (red) for $\epsilon=0.25$ when $x=y=0$ and $\dot{x}=1.14553, \dot{y}=0.829307$, i.e. $K=0.95$. (b,c) Two nearby orbits for $\epsilon = 0.3$, $\alpha = 1/3$ starting at $x=y=0$ with slightly different velocities: (b) $\dot{x}_o = 0.6900848$, $\dot{y}_o = 1.23441602$. In this case we find a Lissajous curve with 4 points on the zero velocity curves and $K = 0.8518518$. (c) $\dot{x}_o = 0.6900856$, $\dot{y}_o = 1.2344156$. In this case we find an  escaping orbit with $K = 0.8518524$.} \label{fig:lissesc}
\end{figure}

Now we will describe the orbits on the $(x, y)$ plane. The orbits cannot cross the CZV ($\dot{x} = \dot{y} = 0$), which has the form
\begin{equation}
    y^2=\frac{2-x^2-\frac{2\epsilon}{3}x^3}{1+2\epsilon x}.
\end{equation}
The value of $y^2$ tends to $\pm\infty$ if $x$ tends to $x=-\frac{1}{2\epsilon}$.
If $\epsilon$ is smaller than $\epsilon_{esc}'$ (e.g. for $\epsilon=0.25$) this value of $x=-\frac{1}{2\epsilon}$ is between the roots $x_2$ and $x_3$ of Eq.(42) in \figref{lissesc}a. Then the curve of zero velocity has two parts. The left part extends to $y=\pm \infty$ along $y$, while the right part is a closed oval curve between $x_1=1.284$ and $x_2=-1.663$ (\figref{figlabel}a). All the orbits starting inside this oval are Lissajous figures with boundaries parallel to the diagonals $x\mp y=0$ (\figref{lissesc}a). The orbits starting on the left of the left CZV escape to $x=-\infty, y= \pm \infty$.

However when $\epsilon>\epsilon_{esc}'$ the situation is more complicated. If we start an orbit for $\epsilon=0.3$ on an invariant curve with small $K$ ($y=0$) (\figref{figlabel}c) we find a Lissajous figure with boundaries parallel to the diagonals $y=\pm x$, like \figref{lissesc}b. The intersections of this orbit with the $y=0$ axis are points ($x, \dot{x}$) on the invariant curve for this $K$. Any orbit starting at any intersection point ($x,\dot{x}$) with $y=0$ and $\dot{y}$ derived from the equation $\dot{x}\dot{y}=K$ has the same boundaries,
which are given by the equation \cite{contopoulos2024}
\begin{equation}
    (x\mp y)^2+\frac{2\epsilon}{3}\left(x\mp y\right)^3=2\mp 2K.\label{eqxy}
\end{equation}

We consider now the solutions of this equation for $x-y$ with given values of $K$. There are two roots of \eqref{eqxy}, depending on the values of $2 - 2K$. The corresponding lines are only approximately at equal distances from the diagonal $x-y=0$ (the third solution of \eqref{eqxy} is far away and it is irrelevant).

The two $x-y$ lines intersect the zero velocity curve at 3 points each. The middle and the upper points of intersection are the corners of the Lissajous figure. However, as $K$ increases from 0 and reaches a certain value $K_0$, the two lower points of intersection of a line $x-y$ with the CZV join and the line becomes a tangent to the CZV. Beyond that value $K>K_0$, the two lower points of intersection do not exist any more and the boundary curve extends to $-\infty$.  Then the orbits are not limited downwards by a straight line parallel to the axis $x+y=0$ but escape to $-\infty$.

In fact, the two boundaries are not exactly symmetric with respect to the line $x-y=0$. Nevertheless, they become tangent to the two parts of the CZV for the same value of $K_0 = 0.851852$ (\figref{lissesc}b).

The transition from Lissajous to an escape orbit is quite abrupt, as shown in \figref{lissesc}b,c. The two curves have the same initial position $x_o = y_o = 0$, but the $\dot{x}_o$ components of the velocity are slightly different. The first is slightly below the red curve of \figref{figlabel}c (for $x=0$) and the second is slightly above it. Then the $\dot{y}_o$, $y_o$ are found from the energy equation \eqref{ham}. The corresponding values of $K$ are found from the equation $K = \dot{x}\dot{y}$ and they are only slightly different. However the two orbits of\figref{lissesc}b,c are totally different, although they are initially very close to each other.

Similar separations between Lissajous and escaping orbits appear along the direction of the other diagonal $x+y = 0$, and for negative values of $K$.

The value of $K_0$ increases as $\epsilon$ increases. When $\epsilon$ goes beyond the escape value $\epsilon_\text{esc}=0.408$, the left and the central parts of the surface of section join (\figref{figlabel}b) and all the orbits become escaping. The corresponding transition value of $K_0$ goes to 0 and to negative values.

In the escaping regions of \figref{figlabel}b,c (yellow regions), we see some invariant curves (green) for fixed values of $K$ larger than $K_0$. These contain the initial conditions of different escaping orbits but with the same $K$.

If instead of the integral $Q_0^2$, we use as zero order integral the integral  generated in the previous section by eliminating the secular terms of order $\epsilon^2$, we find for $\phi=\phi_0+\epsilon\phi_1+\epsilon^2\phi_2+\dots$, where 
$\phi_0=C_0+c_1(2\Phi_0)^2+c_2(2\Phi_{1,0}(2\Phi_{2,0}))c_3(2\Phi_{2,0})^2$ with $c_1=c_3=-1/2, c_2=0$, the expression for the invariant curves gives
\begin{equation}
\phi_0=2\left(-\dot{x}^4+(2-x)^2\dot{x}^2\right),\quad \phi_1=-\frac{4}{3}\epsilon x^2\dot{x}^2,
\end{equation}
i.e. the same terms of Eq.~42 (multiplied by $2$). This is the first order approximation of the integral $\phi$ we have the exact integral. However the formal integral $\phi$ has also a term
\begin{equation}
\epsilon^2\phi_2=\frac{\epsilon^2}{9}\left(5\dot{x}^4+(7x^2-10)\dot{x}^2+2x^4-4x^2+5)(\dot{x}^2+x^2-1)\right),
\end{equation}
which produces different forms of the invariant curves (\figref{figlabel}d for $\epsilon=0.3$).

It is remarkable that the two very different forms of the zero order integrals $\phi_{10}$ and $Q_0^2$ generate the same terms of zero and first order in $\epsilon$ after elimination of the secular terms in $\phi_2$, although they differ in order $\epsilon^2\phi_2$. Of course the differences are very small for small values of $\epsilon$. But for larger $\epsilon$ they become large. Moreover, for $\epsilon > \epsilon_\text{esc}$ (e.g. \figref{figlabel}b) all the orbits inside the boundary escape to infinity. Outside this boundary, no orbits exist because $\dot{y}^2 < 0$.

\subsection{Case $\alpha=0$}

This case has been partly considered already in \cite{ContopMouts1966} with the perturbation $H_1=-xy^2$.

In order to compare it with the above cases we take an energy $E=1$ and $\epsilon$ instead of $-\epsilon$ of the previous paper. Then we have the zero order integrals $\Phi_{1,0}, \Phi_{2,0}$ and $C_0$ and the first order terms are 
\begin{eqnarray}
&\epsilon \Phi_{1,1}=\frac{\epsilon}{3}(xy^2+2x\dot{y}^2-2y\dot{y}\dot{x}),\\&
\epsilon\Phi_{2,0}=\epsilon(xy^2-\Phi_{1,1}),
\end{eqnarray}
while 
\begin{equation}
\begin{aligned}
C_1 =& -\frac{2}{3}\Big[2x^3y^2 - 6\left(x\dot{x}^2y^2 - x\dot{x}^2\dot{y}^2+\dot{x}^3y\dot{y}^2\right) + 2x^3\dot{y}^2+6x^2\dot{x}y\dot{y} + xy^4\\
&- 3xy^2\dot{y}^2-2x\dot{y}^4+ \dot{x}y\dot{y}^3+4\dot{x}y^3\dot{y}\Big]
\end{aligned}
\end{equation}
is given by Eq.~\eqref{C1} with $\alpha=0$.
This is the same as Eq.~(14) of \cite{ContopMouts1966}, but with $\-\epsilon$ instead of $\epsilon$.

In this case we find secular terms in $\Phi_{1,2}, \Phi_{2,2}$ and $C_2$ and in order to avoid them we use an integral with zero order terms as in Eq.~\eqref{comb} with
 $c_1=1/6, c_2=0, c_3=1/12$, as in 
 \cite{ContopMouts1966}.

\begin{figure}[H]
\centering
\includegraphics[scale=0.3]{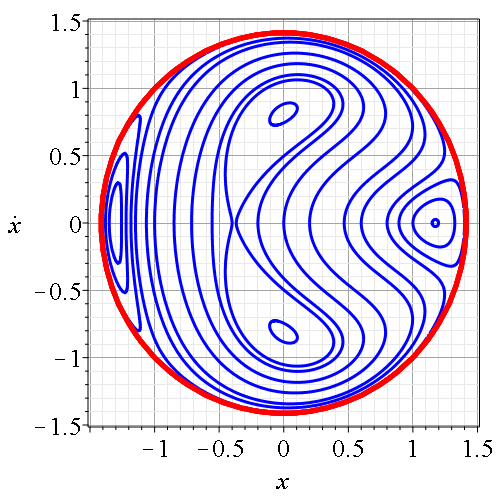}
\includegraphics[scale=0.45]{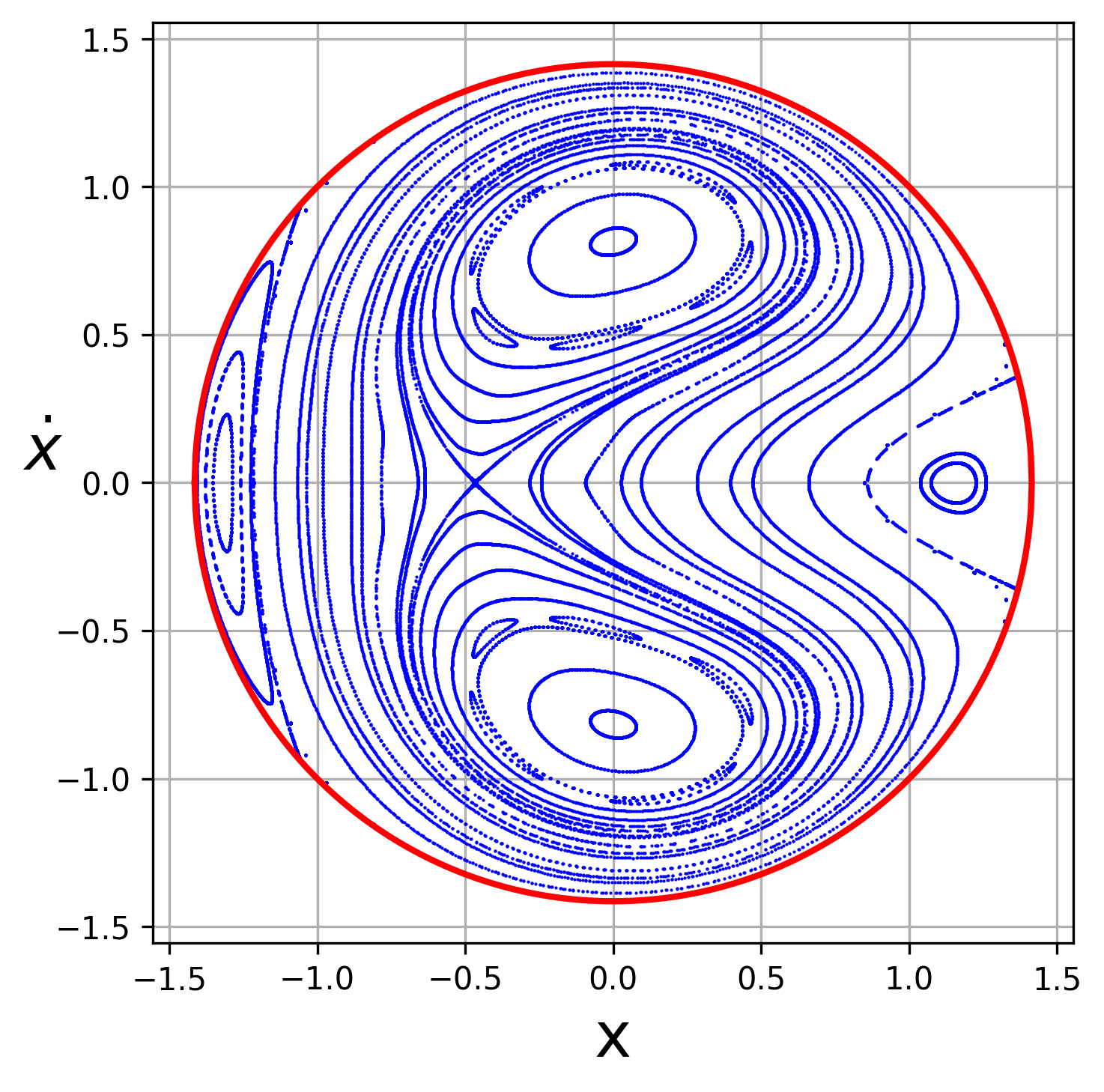}
\includegraphics[scale=0.45]{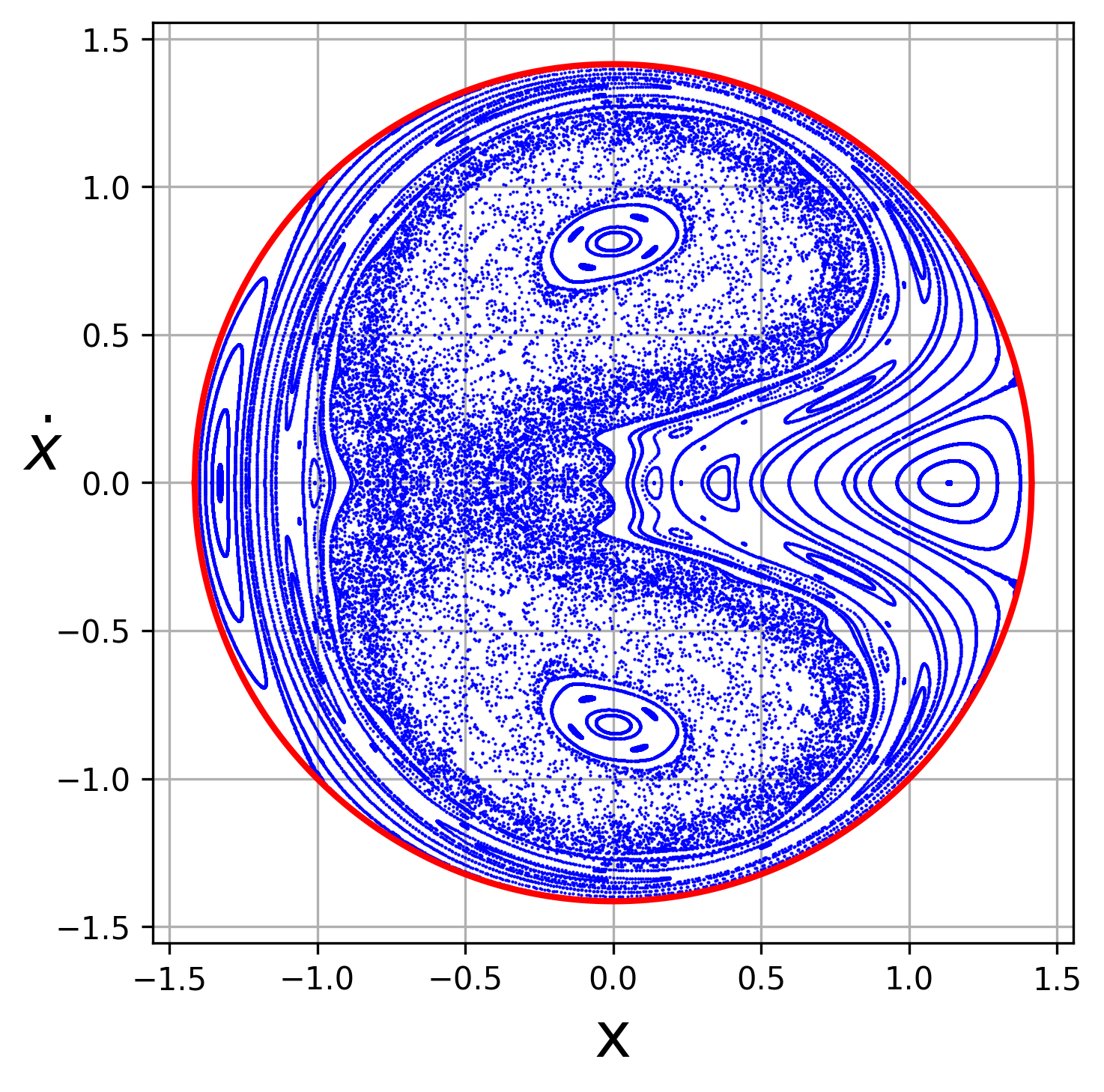}
\includegraphics[scale=0.3]{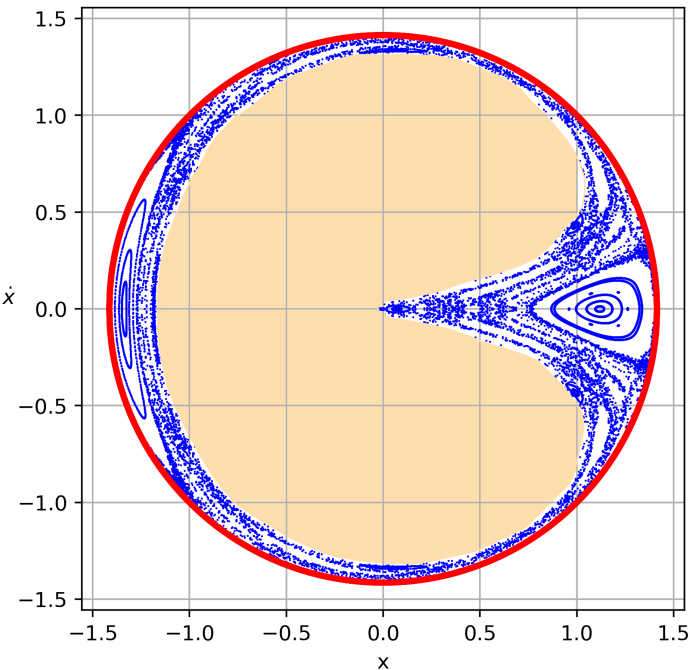}
\caption{Theoretical (a) and exact (b, c, d) surfaces of section in the cases with $\alpha = 0$ and (a, b) $\epsilon = 0.3$, (c) $\epsilon = 0.35$ and (d) $\epsilon = 0.4$. In the last case the yellow color indicates escaping orbits along the $y$-direction.}\label{fig:alphazero}
\end{figure}

%Using this integral up to the first order in $\epsilon$ and taking a surface of section $y=0$ we find 
%\begin{equation}
%\Phi=.....+\epsilon()
%\end{equation}
Then we calculate several invariant curves on a surface of section $y=0$ for various values of $\epsilon$ (\figref{alphazero}).
Thus we have an integral
\begin{equation}
\phi=\phi_0+\epsilon\phi_1+\epsilon^2\phi_2
\end{equation}
without secular terms up to order $\epsilon^2$ (and $\epsilon$) where
\begin{equation}
\phi_0=\frac{1}{3}(2\Phi_{1,0})^2+\frac{1}{12}(2\Phi_{2,0})^2+C_0
\end{equation}

The present invariant curves are different from those of \cite{ContopMouts1966} because we used there different $\alpha$ and energy and a single value of $\epsilon$. In  \cite{ContopMouts1966} we calculated the invariant curves by establishing the integral up to order $8$, in order to show the accuracy of the formal integral $\Phi$. However we considered only one value of $\epsilon$ and we did not study the chaotic orbits. Here we consider only a second order approximation of this integral but we calculate the invariant curves for various values of $\epsilon$ and we compare these with the exact numerical invariant curves.

In \cite{ContopMouts1966} we made numerical calculation for a small value of $\epsilon$ up to order $\epsilon^8$. In those cases there is no appreciable chaos.  However, for larger $\epsilon$ we find the onset of chaos as in the non integrable H\'{e}non-Heiles system. In particular, when $\epsilon=0.3$ ($\alpha = 0$) the integral $\Phi$ gives only the main invariant curves (\figref{alphazero}a) while the chaotic regions in the exact problem have very little chaos (\figref{alphazero}b). In \figref{alphazero}b we see a number of small islands of SPOs, and this is an indication of the onset of chaos. This figure (\figref{alphazero}b) is, in general, very similar to the theoretical figure (\figref{alphazero}a). On the other hand, for $\epsilon=0.35$, we see a lot of chaos around a UPO close to the origin (\figref{alphazero}c). Finally, for $\epsilon=0.4$ (\figref{alphazero}d) most orbits escape to infinity along the $y$ direction (yellow region). This is due to the fact that the value $\epsilon = 0.4$ is larger than the escape value $\epsilon_\text{esc}^\prime = \frac{1}{\sqrt{8}} = 0.354$ (for $\alpha = 0$). Then only two islands remain on the left and on the right. Therefore, for $\alpha = 0$,  the third integral theory is applicable at least up to $\epsilon=0.3$ but breaks down for $\epsilon=0.35$.

Consequently, the third integral in the non integrable H\'{e}non-Heiles case when we have two perturbation terms ($\epsilon (xy^2-\frac{x^3}{3}$) is not applicable for values of $\epsilon$ between $0.1$ and $0.3$, while it is applicable in the case of a single perturbation term ($\epsilon xy^2$). This difference reminds us to be very careful in applying the third integral in various cases.

All these considerations refer to the approximation of the third integral truncated at various orders in $\epsilon$ ($\epsilon^2$  in the numerical examples above) and do not have any chaos. However, in the exact numerical cases there are chaotic regions, except in the  integrable case $\alpha=1/3$  and, of course, the zero order approximation of the integrals. Chaos is very important for large $\epsilon$, i.e. when $\epsilon$ approaches the escape perturbation.

\subsection{Surfaces of section for $\alpha > 0$}

We start with the case $\alpha=0$ (\figref{eps0}b) that has two large islands of stability on the left and on the right, and two large islands above and below the origin, separated by a UPO at the origin.

% If the perturbation $\epsilon$ is small, the surface of section for $\alpha\leq 0$ is similar to \figref{eps0}b, but the unstable orbit is on the left of the origin. As $\alpha$ decreases, the islands on the left and on the right increase, while the islands above and below the origin decrease becoming very small as $\alpha$ approaches the value  to $\alpha=-1/3$. (\figref{005minus1over3}b) (non integrable H\'{e}non-Heiles system for $\epsilon=0.05$). As $\alpha$ becomes smaller than $-1/3$, the unstable periodic orbit on the left of the origin becomes stable and a small island elongated along is formed around it. The transition from instability to stability occurs for $\alpha$ little smaller than $\alpha=-1/3$.

\begin{figure}[H]
\centering
\includegraphics[scale=0.3]{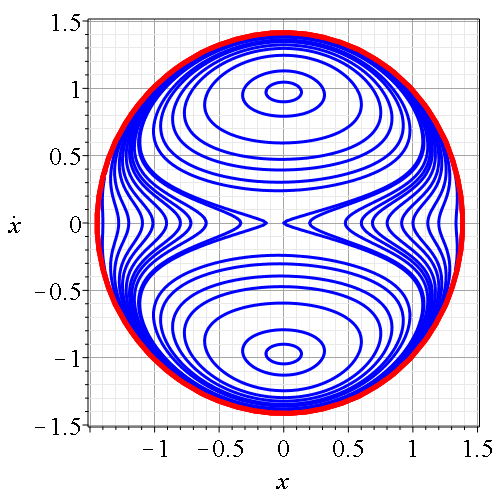}
\includegraphics[scale=0.42]{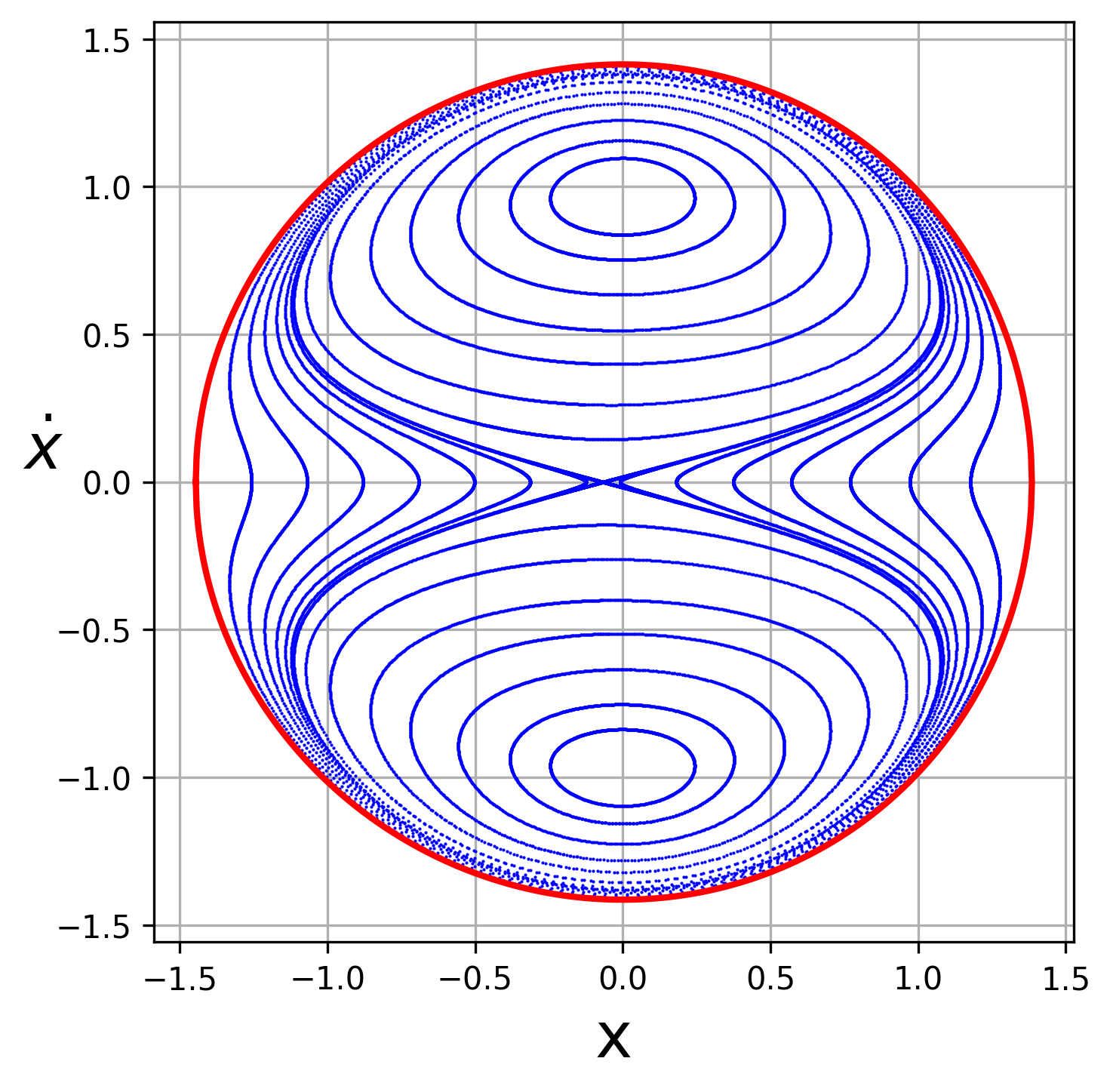}
\caption{Theoretical (a) and exact (b) surfaces of section of the case with $\epsilon = 0.05$ and $\alpha = 0.3$.}\label{fig:00503}
\end{figure}

\begin{figure}[H]
\centering
\includegraphics[scale=0.3]{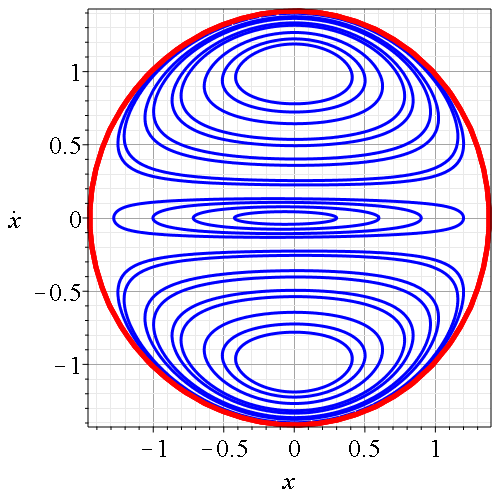}
\includegraphics[scale=0.42]{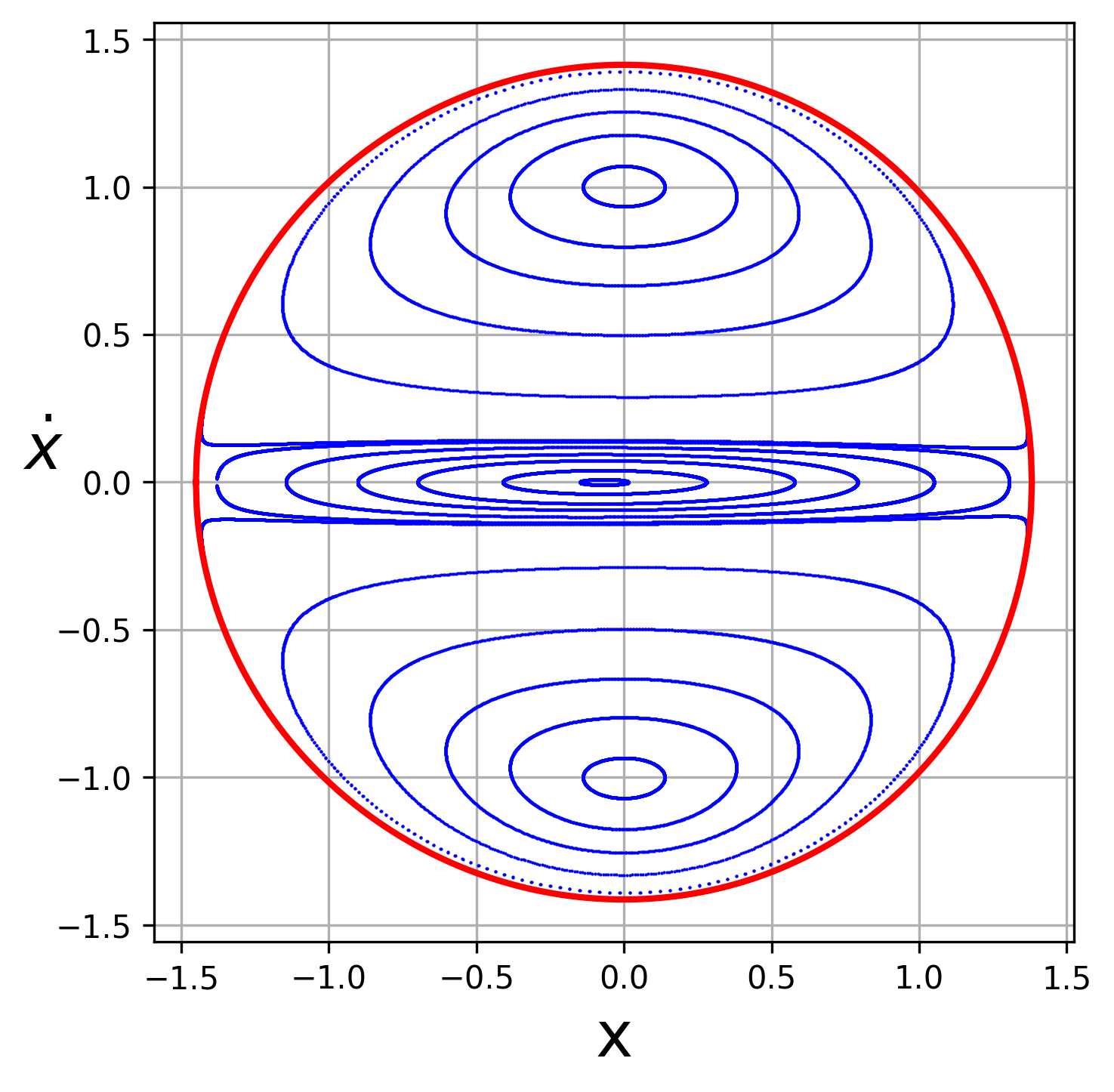}
\caption{Theoretical (a) and exact (b) surfaces of section of the case with $\epsilon = 0.05$ and $\alpha = 0.34$.}\label{fig:005034}
\end{figure}

When $\alpha$ increases from 0, the islands above and below the origin become larger (\figref{00503}), while those on the left and on the right become smaller and disappear at the edges of the surface of section when $\alpha = \alpha_d = 0.133$. The exact value is about the same for $\epsilon = 0.05$ and $\epsilon = 0.3$. The theoretical value is $\alpha_d$=0.133 for $\epsilon = 0.05$, but it is $0.118$ for $\epsilon = 0.3$. This deviation is due to the small truncation order of the third integral. Moreover, the theoretical invariant curves for $\epsilon=0.05$ ($\alpha=0.3$) (\figref{00503}a) are very similar to the exact invariant curves (\figref{00503}b).

\par Between $\alpha = 0$ and $\alpha = 1/3$ there is a UPO on the x-axis and its asymptotic curves surround both the upper and lower islands (\figref{00503}a,b) which become larger as $\alpha$ increases.
\par As $\alpha$ reaches the value $\alpha = 1/3$, the UPO (for $\epsilon=0.3, \dot{x}=0$) tends to $x = -0.52$ and the invariant curves surrounding both upper and lower islands disappear. In this case there are separate invariant curves around the points ($x=0, \dot{x}=\pm 1$) above and below the $x$-axis for every $\epsilon < \epsilon_\text{esc} = \frac{1}{\sqrt{6}} = 0.408$. If $\epsilon$ is larger than $\epsilon_\text{esc}^\prime = 0.408$ then we have escapes along the y-axis for orbits starting close to ($x=0, \dot{x} = \pm 1$) (\figref{figlabel}c).
\par When $\alpha$ goes beyond the value $\alpha = 1/3$, there is a stable periodic orbit close to the origin on the $x$-axis, surrounded by an island with invariant curves elongated along the $x$-axis (\figref{005034}a (theoretical) and (b) numerical). In the limit $\alpha \rightarrow 1/3^+$, this stable periodic orbit is again at $x=-0.52$ (for $\epsilon=0.3,\dot{x} = 0$). 
\par The outermost curve of the central island reaches the boundary of \figref{005034}b  at two points on the left and two points on the right. These points represent UPOs on the boundary of \figref{005034}b (red curve). 

\begin{figure}[H]
\centering
\includegraphics[scale=0.295]{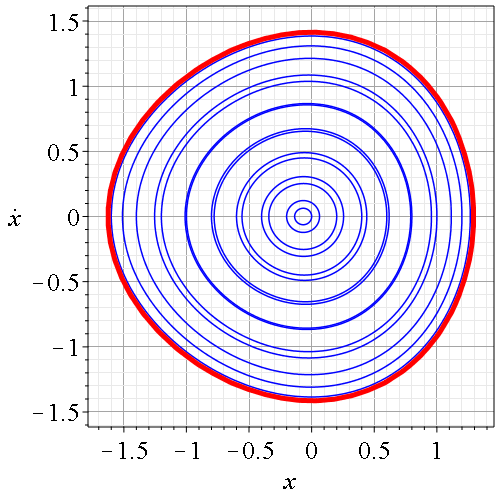}
\includegraphics[scale=0.42]{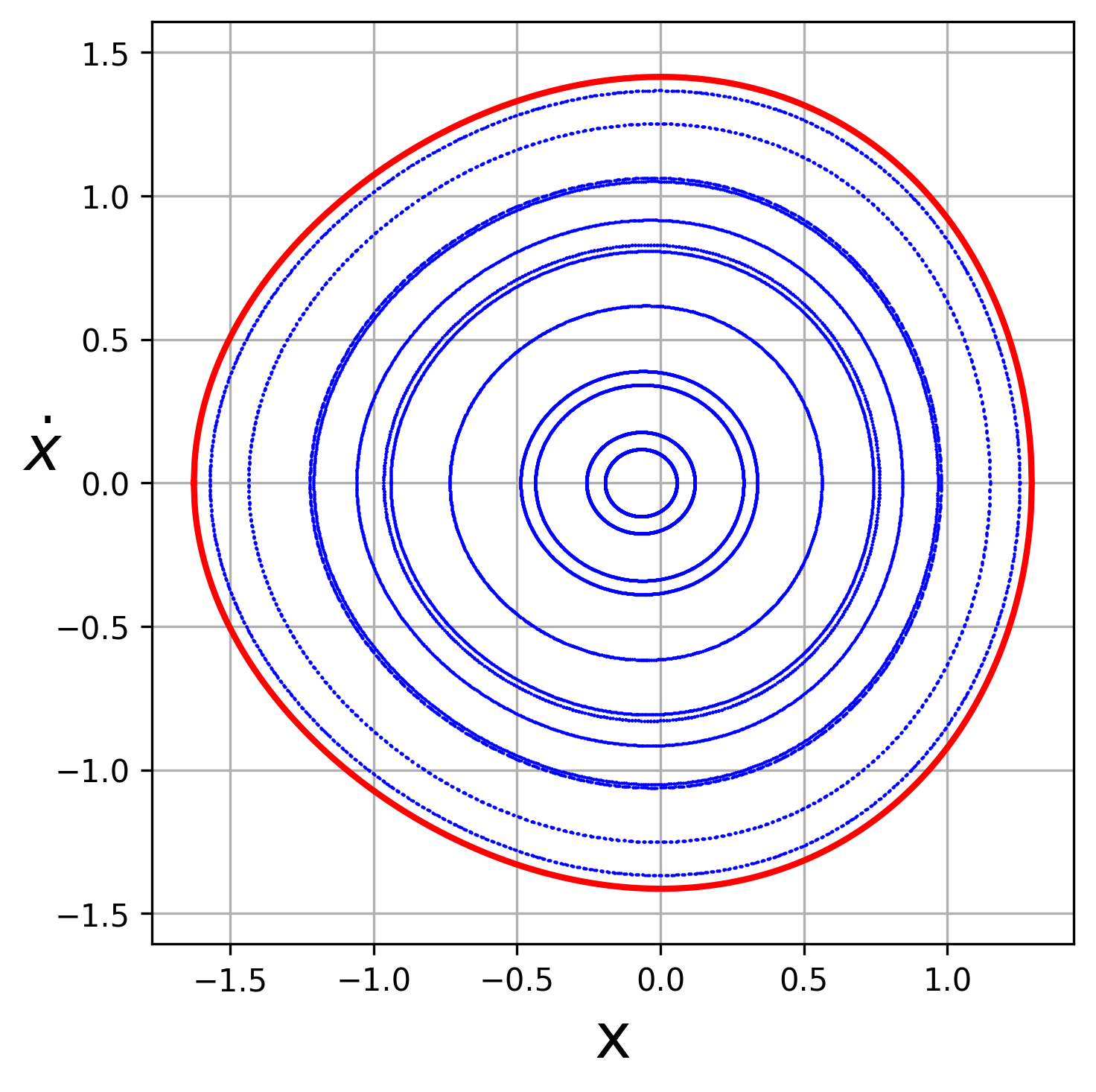}
\caption{Theoretical (a) and exact (b) surfaces of section of the case with $\epsilon = 0.05$ and $\alpha = 1.5$.}\label{fig:00527}
\end{figure}

As $\alpha$ increases with constant $\epsilon=0.05$ the central island becomes larger while the upper and lower islands become smaller and beyond $\alpha=2/3$ they disappear. For $\alpha>2/3$ the central island covers the whole available space $(x,\dot{x})$ around the central periodic orbits (Fig.~\ref{fig:00527}). The theoretical (\ref{fig:00527}a) and the exact (\ref{fig:00527}b) invariant curves agree with each other. No appreciable chaos exists in the exact case. This situation exists all the way up to $\alpha=2.722$ (for $\epsilon=0.05$).

% As $\alpha$ approaches the escape value $\alpha_{esc}=2.722$ (\figref{00527}a,b) the limiting curve (red) becomes elongated along the negative $x$-axis, but no appreciable chaos appears. In this case again the theoretical invariant curves (\figref{00527}a) are similar to the exact ones (\figref{00527}b).

\begin{figure}[H]
\centering
\includegraphics[scale=0.31]{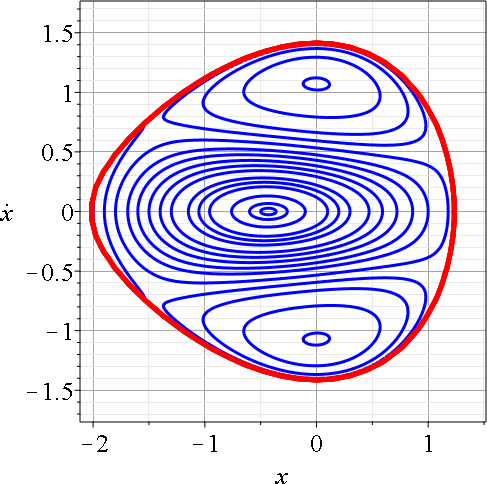}
\includegraphics[scale=0.3]{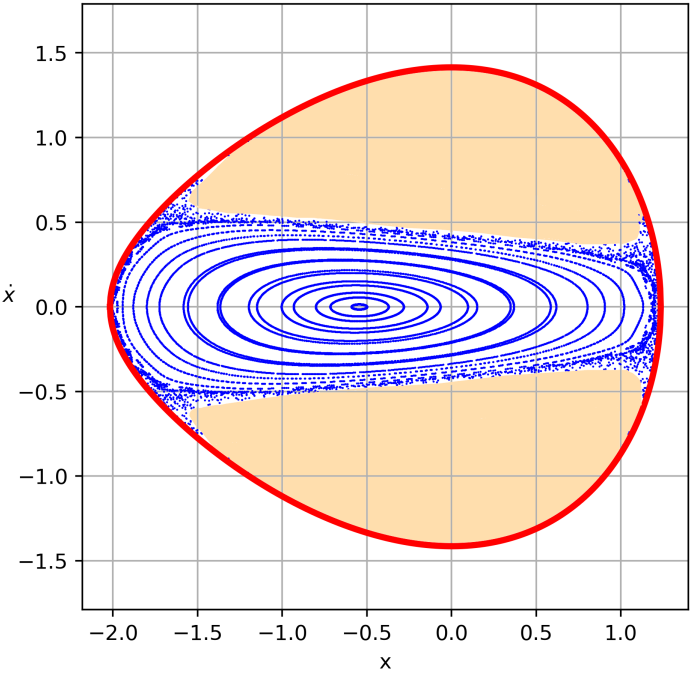}
\caption{Theoretical (a) and exact (b) surfaces of section for $\epsilon = 0.3$, $\alpha = 0.42$. The yellow regions contain initial conditions of orbits escaping towards $y = \pm \infty$. Chaos appears in the exact case (b) near the boundaries of the central island.}\label{fig:temp1}
\end{figure}

As $\epsilon$ increases from $\epsilon=0.05$ the central island becomes larger than that in \figref{005034}a,b. E.g. in the cases $\epsilon=0.3$ we see a large central island (\figref{temp1}a,b) and some chaos appears near the four UPOs on the intersections of the boundary of the central island with the boundary of the surface of section that corresponds to $\dot{y}^2=0$
\figref{temp1}b. In this case where $\alpha$ is larger than the escape perturbation $\alpha'_{esc}=0.280$, we have two large domains of orbits escaping along the $y$-direction above and below the central island (yellow regions). The theoretical central island (\figref{temp1}a) is similar to the exact (\figref{temp1}b), but the theoretical surface of section does not have any chaos or escapes.

\begin{figure}[H]
\centering
\includegraphics[scale=0.45]{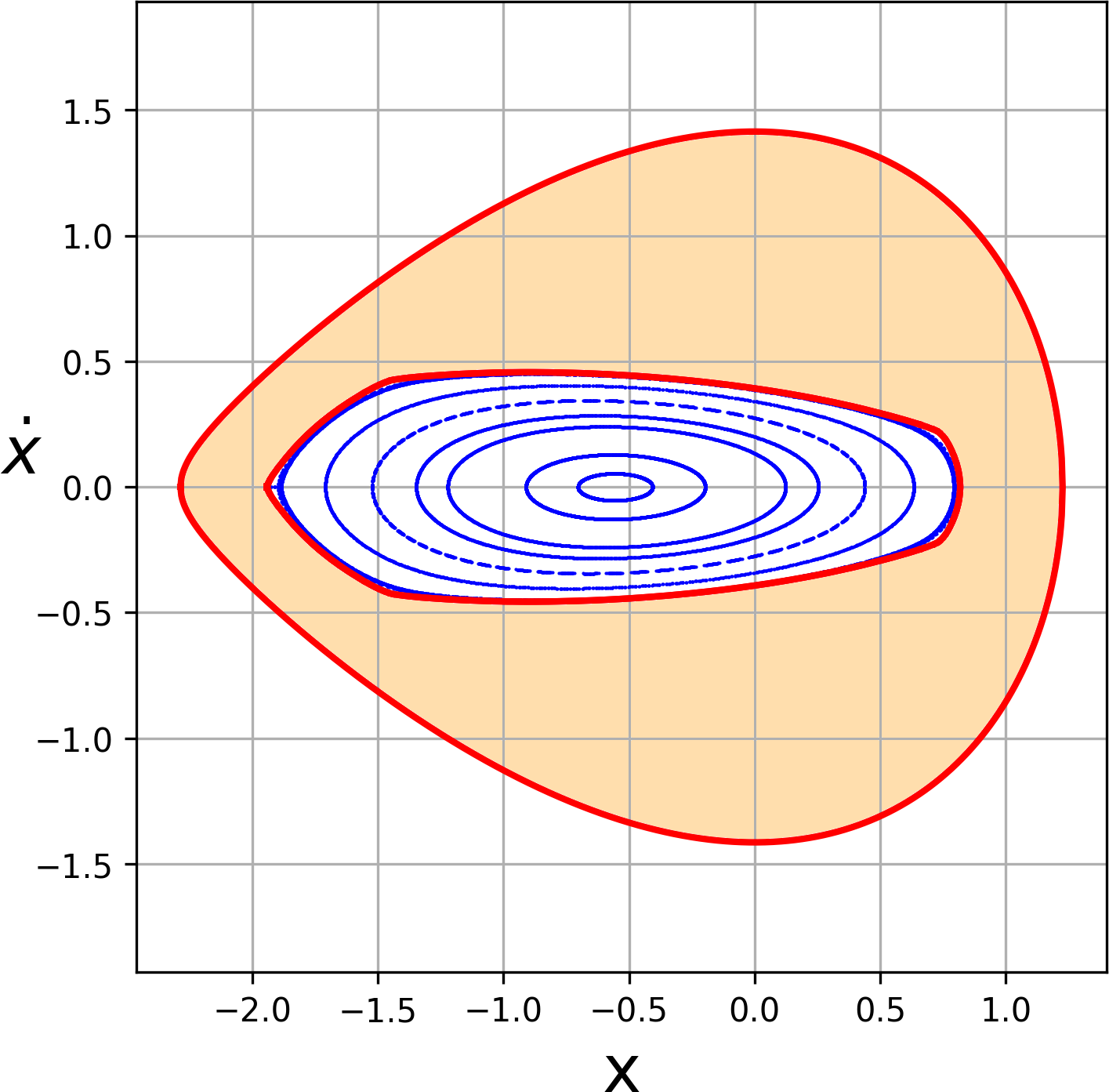}
\includegraphics[scale=0.45]{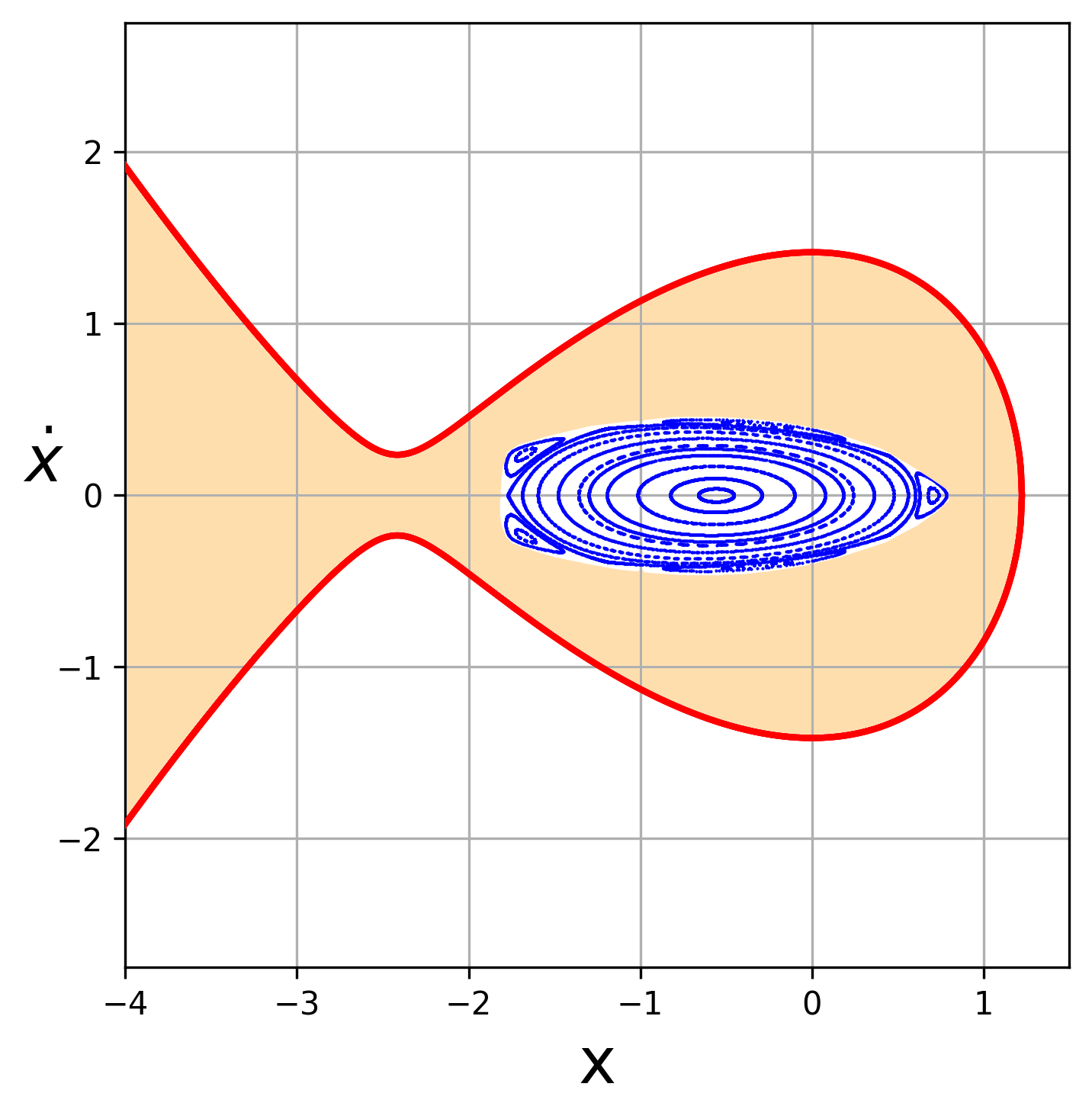}
\caption{Exact surfaces of section for $\epsilon = 0.3$ and: (a) $\alpha = 0.45$ and (b) $\alpha = 0.46$. The yellow regions contain initial conditions of orbits escaping towards $y = \pm \infty, x=-\infty$.}\label{fig:temp}
\end{figure}

As $\alpha$ increases further and approaches the escape value $\alpha_{esc}=0.454$ (for $\epsilon=0.3$) the escapes towards $y=\pm\infty$ cover most of the available space (\figref{temp}a). However a central island continues to exist even beyond the escape perturbation (\figref{temp}b).

For even larger $\alpha$ ($\alpha=2$) we have another integrable case \cite{lakshmanan1993painleve,lakshmanan2012nonlinear} but we are not to going discuss it in the present paper.

% For $\epsilon = 0.3$ and $\alpha = 0.42$ (i.e. large $\alpha$, close to the escape value $\alpha_\text{esc} = 0.454$) we see chaos around them (\figref{temp}b). In this case most orbits around above and below the central island  escape to infinity. The boundaries of the escaping regions in the case $\epsilon = 0.3$ are very close to the boundaries of the chaotic regions and the theoretical invariant curves (\figref{temp}a) are similar to the exact curves in the central region (\figref{temp}b). But the theoretical surface of section has no chaos and no escapes.

% \par For larger $\alpha$, only a small island remains on the left of the origin in the exact case, and most orbits (including the chaotic ones) above and below it escape to infinity.

% When $\alpha=1/3$ we have the integrable H\'{e}non-Heiles case where the surface of section is occupied by the two large islands above and below the x-axis. For $\alpha>1/3$ a large island is found around a  stable periodic orbit (a little to the left of the origin), which is elongated along the x-direction (\figref{005034} theoretical and exact invariant curves for $\epsilon = 0.05$, $\alpha = 0.34$).

% In all the cases, except the case $\alpha=1/3$, this disappearance occurs for $\epsilon = 0.05$ and $\alpha \geq 0.133$. There are also infinitely many stable and UPOs of higher order. Around the stable orbit there are small islands of stability, and around the unstable orbits there is chaos. The chaotic domains are extremely small, but they  increase abruptly beyond the critical value of $\epsilon$ (e.g beyond $\epsilon=0.3$ in the cases $\alpha=-1/3$ and $\alpha=0$ ).  

% The formal integral, gives approximately the forms of the exact invariant curves when epsilon is small, but deviates from the exact curves around the unstable periodic orbit close to the origin, even when $\epsilon=0.05$. However, the variations are marked larger. 

\subsection{Limiting Cases}
If we consider a more general perturbation
\begin{equation}
\epsilon(xy^2+\alpha x^3+\beta x^2y+\gamma y^3),
\end{equation}
we find the secular terms of the forms
\[
\begin{aligned}
&(2\Phi_{1,0})^2(2\Phi_{2,0})\sin(2t_0), (2\Phi_{1,0})(2\Phi_{2,0})^2\sin(2t_0)\\
&(2\Phi_{1,0})(2\Phi_{2,0})^{\frac{3}{2}}\sin(3t_0),[(2\Phi_{1,0}^5)(2\Phi_{2,0})]^{\frac{1}{2}}\sin(t_0)
\end{aligned}
\]
with coefficients depending on $\alpha, \beta, \gamma$ and $c_1, c_2, c_3$. However, the 3 relations containing $\sin(t_0)t$ have terms proportional to $\beta$ and $\gamma$, thus they are satisfied only if $\beta=\gamma=0$. The coefficient of the term with $\sin(3t_0)t$ is proportional to $(\gamma-\alpha \beta)$ and therefore it is  satisfied only for $\beta=\gamma=0$. 

The coefficients of the terms with $\sin(2t_0)t$ for $\beta=\gamma=0$ are those with only $\alpha$ and coincide with the coefficients found already in section 2.  As a consequence, for any value of $\alpha$, we can find appropriate values of $c_1, c_2$ and $c_3$, in order to eliminate the secular terms of $\phi_2$. But we cannot eliminate all the secular terms  unless $\beta=\gamma=0$. This fact means that we cannot calculate a non resonant integral if the perturbation (52) contains terms with $\beta$ and $\gamma$. But if the first two terms of (Eq.~52) are 0, then, because of the symmetry, we can compute integrals with perturbation $\epsilon(\beta x^2y +\gamma y^3)$, which have only secular terms with $\sin(2t_0)t$ that can be eliminated.

\section{The onset of chaos}

\begin{figure}[H]
\centering
\includegraphics[scale=0.4]{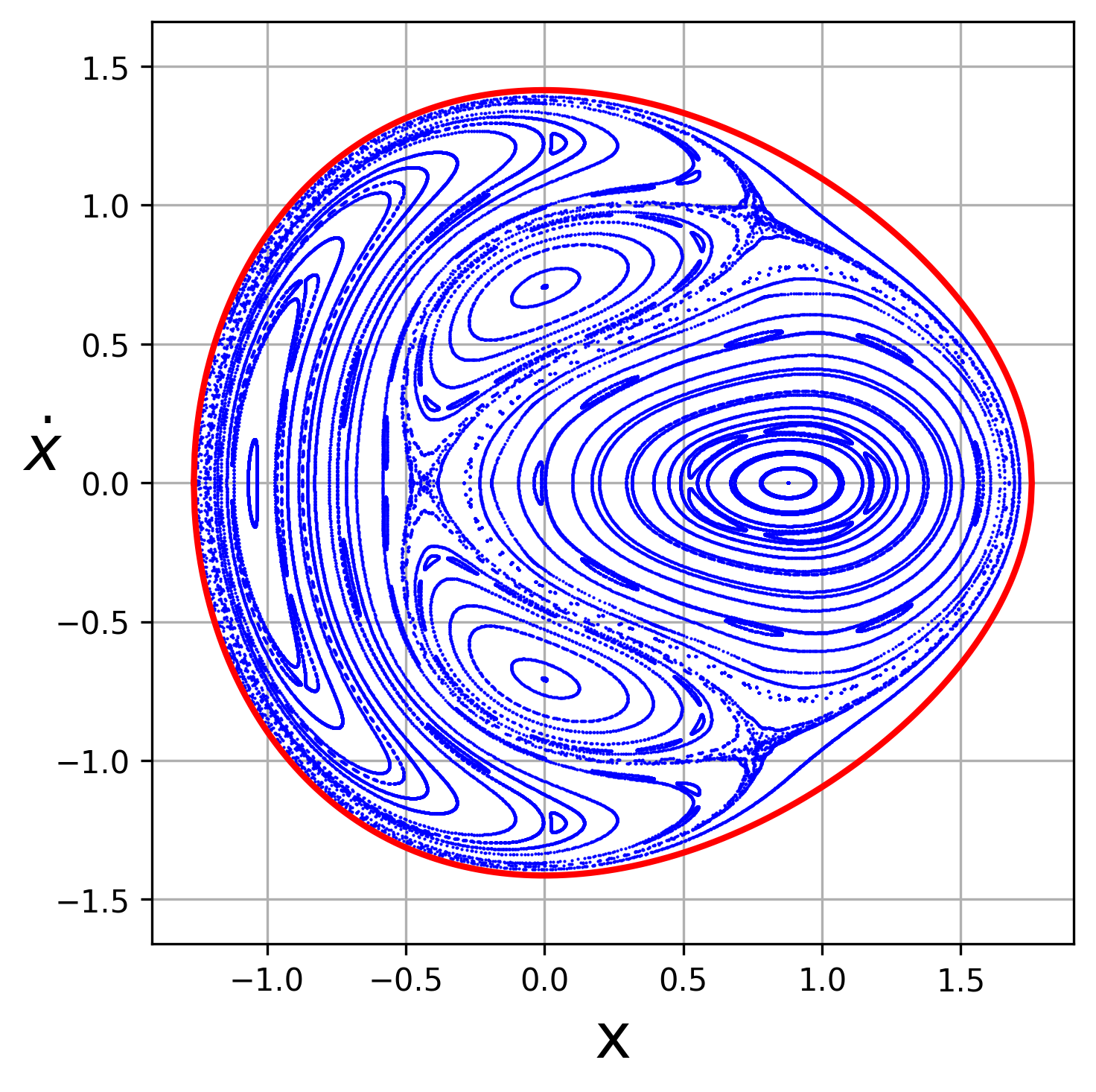}
\includegraphics[scale=0.4]{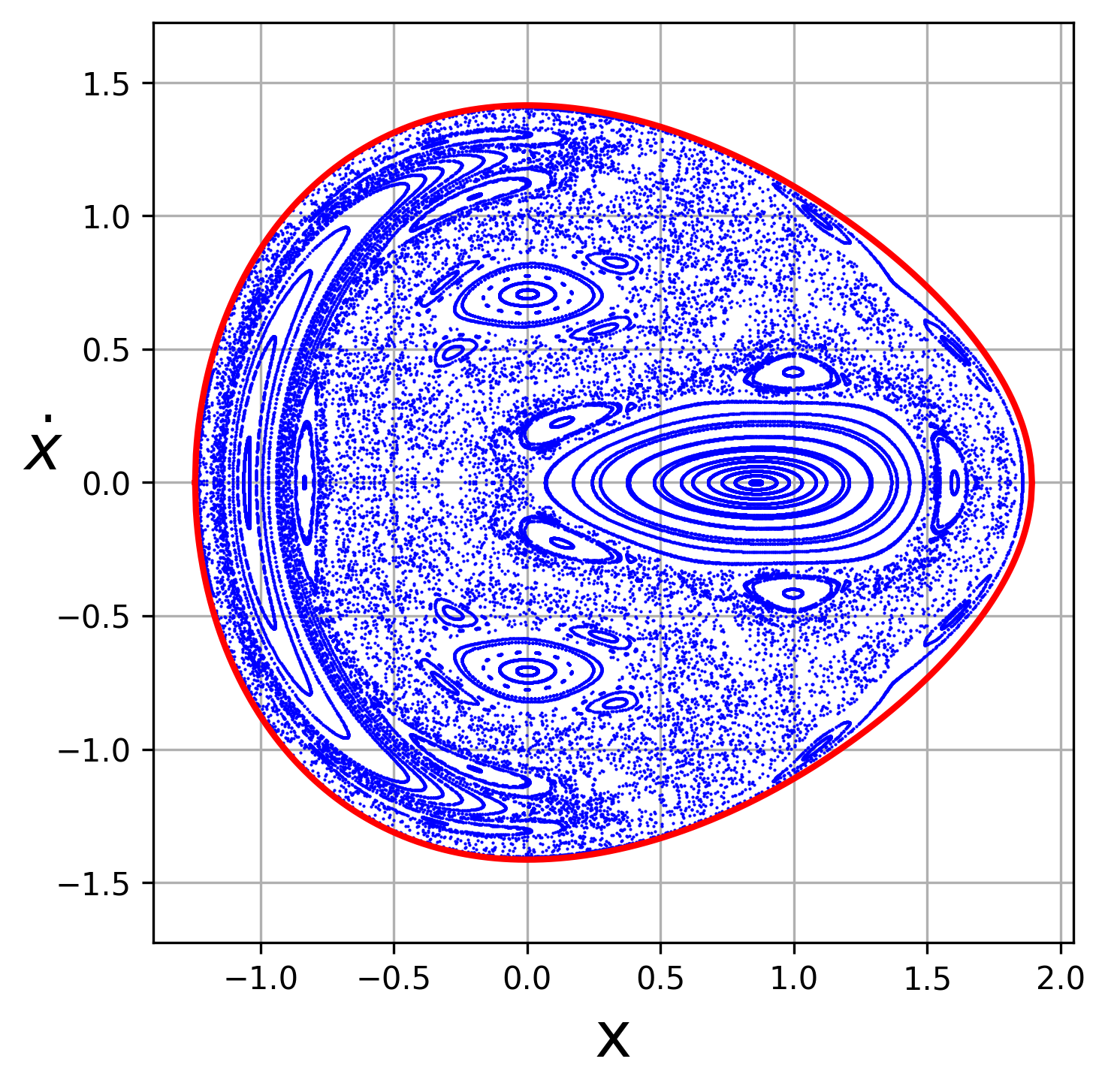}
\caption{Exact surfaces of section in the cases with $\alpha = -1/3$ (non-integrable H\'{e}non-Heiles system) for (a) $\epsilon = 0.3$ and (b) $\epsilon = 0.35$.}\label{fig:minus1over3}
\end{figure}

Chaos appears near UPOs. We consider now the onset of chaos in the non integrable H\'{e}non-Heiles system. In \figref{minus1over3}a, we see small chaotic regions near the UPOs  of period 1 ($x=-0.435$, $\dot{x}=0$) and  ($x=0.755$, $\dot{x}= \pm 0.944$) for $\epsilon = 0.3$ ($\alpha = -1/3$). In \figref{minus1over3}b for $\epsilon = 0.35$, we see a huge chaotic region mainly on the left of the origin, extending to the right above and below the island around the main stable periodic orbit at $x = 0.857$, $\dot{x} = 0$.

In \figref{minus1over3}a,b we see many more small islands of stability. Between them there are UPOs that generate their own small chaotic regions, or contribute to the general chaotic domain.

 One way to locate those periodic orbits around the main stable periodic orbit on the right of the origin is by calculating the rotation numbers of orbits starting on the x-axis along various distances ($x - x_o$) from the main periodic orbits. The rotation number ($RN$) is the average angle of the successive intersections of an orbit with the surface of section as seen from the main periodic point.

\begin{figure}[H]
\centering  
\includegraphics[scale=0.4]{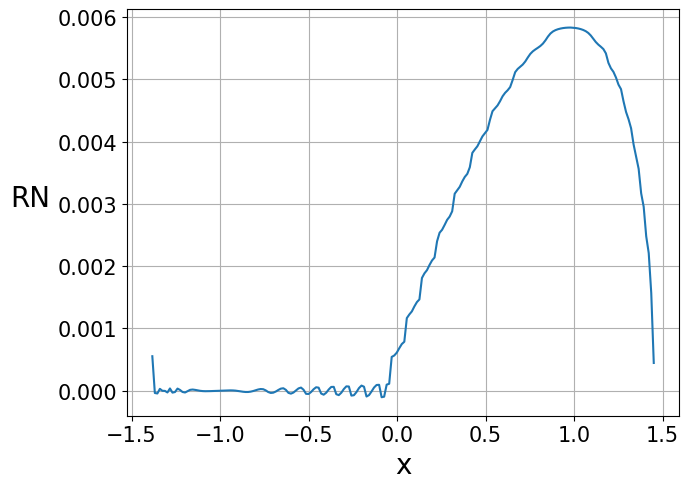}[a]
\includegraphics[scale=0.4]{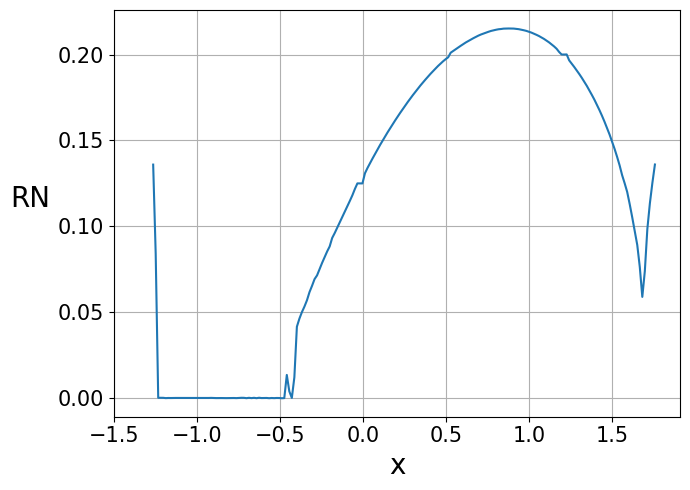}[b]
\includegraphics[scale=0.4]{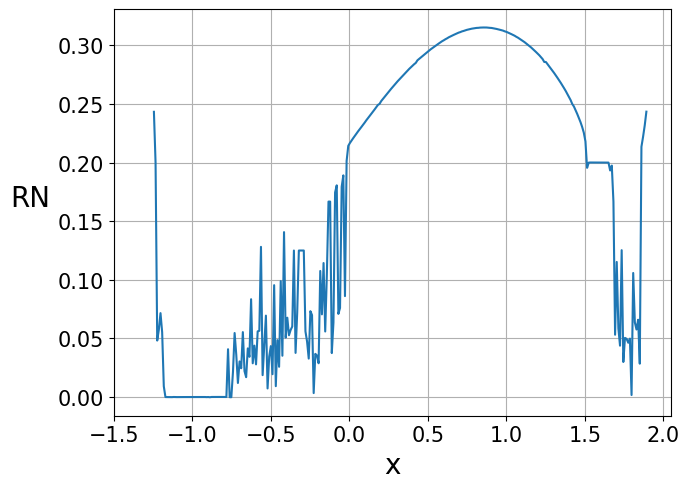}[c]
\caption{The $RN$ as a function of $x$ for $\dot{x} = 0$ in the case $\alpha = -1/3$ around the main stable periodic orbit on the right at (a) $x_o = 0.982$ ($\epsilon = 0.05$),  (b) $x_o = 0.878$ ($\epsilon = 0.3$) and (c) $x_o = 0.857$ ($\epsilon = 0.35$).}\label{fig:rots}
\end{figure}

 For small $\epsilon$ ($\epsilon = 0.05$), the $RN$ varies from 0 to a maximum $RN$ = 0.0058 (\figref{rots}a) at the central periodic orbit ($x_o = 0.582$, $\dot{x} = 0$) and reaches the value $RN = 0$ close to $x = 0$. There are periodic orbits, stable and unstable, at every rational value of $RN$, but their islands and chaotic domains are extremely small.
 
 In \figref{rots}b, we see that the maximum $RN$ for $\epsilon = 0.3$, close to $RN = 0.21$ is at the right periodic point itself. A little on the right of the maximum we see one of a set of 5 islands with rotation number $RN = 0.2$. On the left of the central point we have at $RN=0.2$, an unstable point of a period-5 UPO. Further to the left and to the right we see two small straight line segments with  $RN= 0.125$ that represent 8 islands around a stable period-8 orbit. Finally, on the left of $x=-0.5$ we see a plateau at $RN=0$, representing the curves around the left main periodic orbit at $x=-1.015$. At $x=-0.435$ we have the main UPO and around this point we see some irregular variations of $RN$, representing the chaotic orbits close to this UPO.

 Similar phenomena are found if we calculate the $RNs$ around the left stable periodic orbit of \figref{temp}a. Near the left and the right borders of the surface of section, the $RN$ increases again, corresponding to closed invariant curves near the boundary of \figref{rots}a, which represents the periodic orbit $y = \dot{y} = 0$.

 These phenomena are magnified for $\epsilon = 0.35$ (\figref{rots}c) giving the corresponding $RNs$ around the right stable periodic orbit. In this case the maximum $RN$ at the stable periodic orbit is  $RN\simeq 0.331$. At a certain distance on the right we see a small plateau at  $RN=0.2$ corresponding to one of the 5 islands around a stable periodic orbit of period 5. On the left of the maximum at  $RN=0.2$ there is an unstable point (out of 5 unstable points around the stable center) and on the left of this point we see large irregular variations of $RN$, representing the chaotic orbits of \figref{minus1over3}b. Further to the left there is a plateau at $RN=0$ representing the island of orbits around the left stable periodic orbit at $x=-0.9$. At the limits on the left and the right we see again chaotic regions.
 
We note that in the chaotic regions there are several UPOs (in fact, an infinite number of them). From every unstable point emanate two stable and two unstable asymptotic curves (in opposite directions) that intersect. The intersections of the asymptotic curves of the same periodic orbit are `homoclinic points',  while the intersections from different periodic orbits are `heteroclinic points'. The existence of heteroclinic points is the main characteristic of chaos, because these orbits (and nearby orbits) approach the regions of different periodic orbits as $t\rightarrow \infty$ and $x\rightarrow -\infty$.

The $RN$  changes abruptly close to an unstable point. In a small interval there are several rational numbers (in fact infinite) that correspond to UPOs of high multiplicity that intersect each other at nearby heteroclinic points. However, a large degree of chaos appears when we have intersections of asymptotic curves of distant UPOs. E.g. in \figref{minus1over3}b we see that in the region $-7.5 < x < 0$ there is a large interval of chaos that includes UPOs of periods 7, 8, etc., besides the main unstable point at $x= -0.5$.

\begin{figure}[H]
\centering
\includegraphics[scale=0.25]{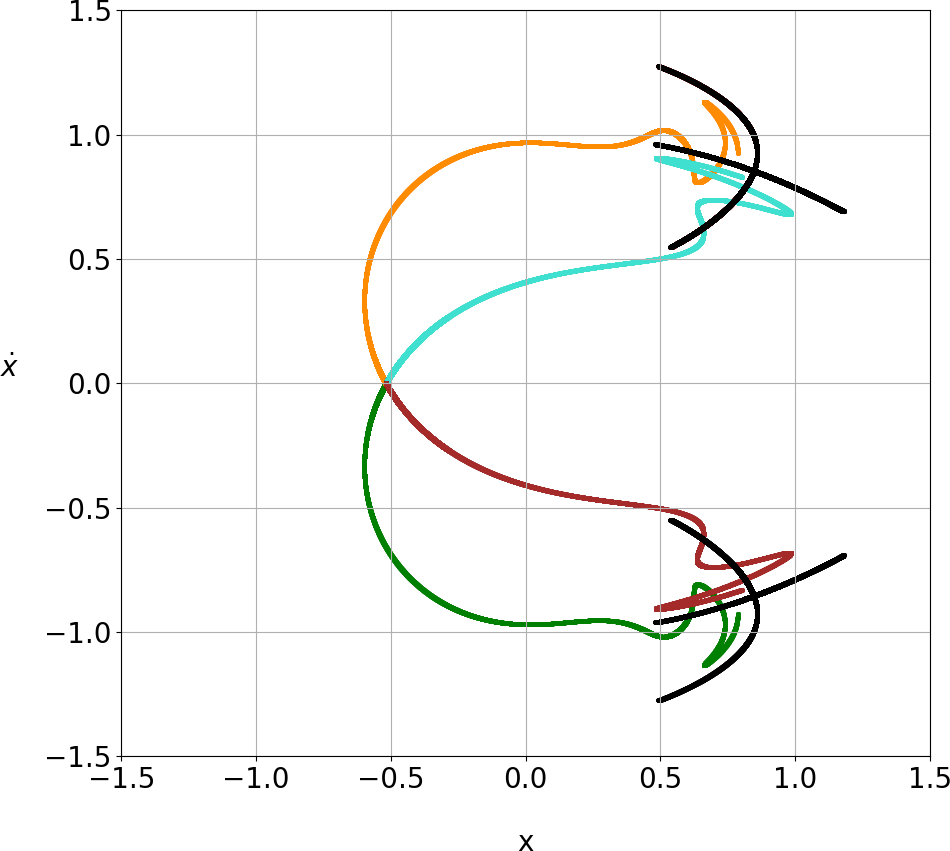}[a]
\includegraphics[scale=0.25]{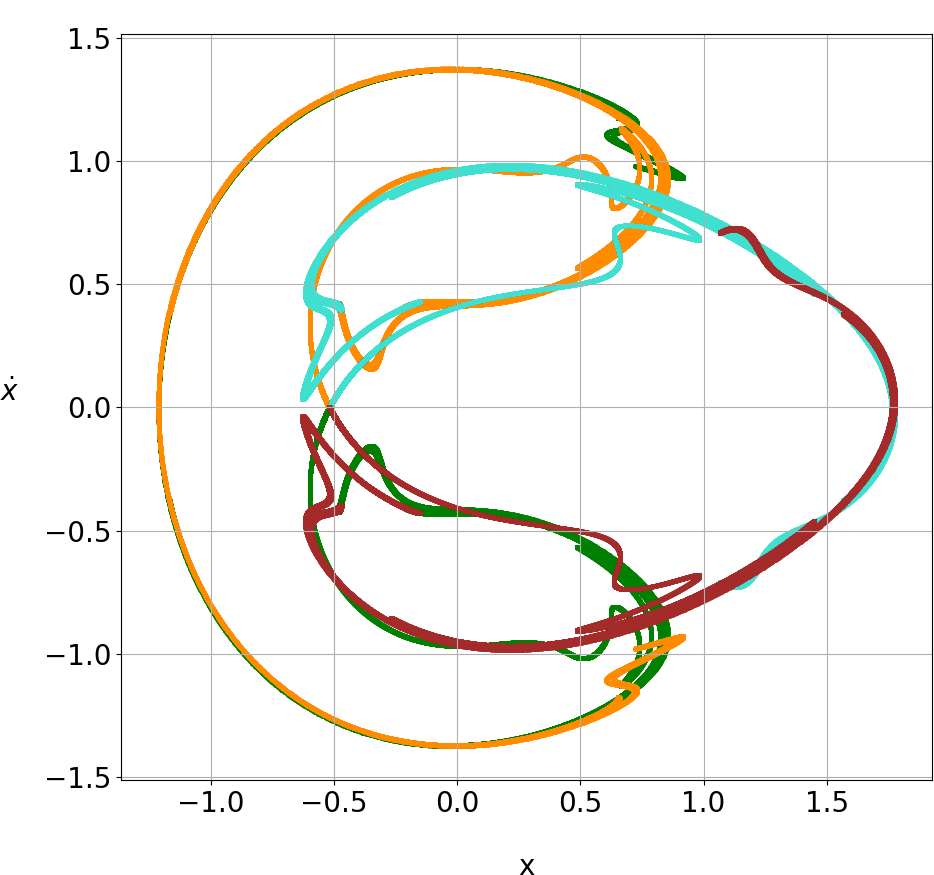}[b]
\caption{(a) The asymptotic curves for $\epsilon = 0.35$, $\alpha = -1/3$ of the main UPO on the x-axis with $x_o = -0.519$. They are formed by 13 images of 20000 initial conditions inside an interval of length $10^{-6}$ in the eigendirections of the periodic orbit. We show also in black color the initial part (10 images) of the asymptotic curves of the period 1 orbits ($x = 0.754$, $\dot{x} = \pm 0.944$)  b) Extension of the asymptotic curves of the central periodic orbit up to 16 images.}\label{fig:asympts1}
\end{figure}

The fact that the asymptotic curves of these orbits intersect each other at heteroclinic points is clearly shown in \figref{asympts1}, where we see the intersection of the asymptotic curves of the central UPO with those of  the unstable period-7 orbit.

 In \figref{minus1over3}a ($\epsilon = 0.3$) the asymptotic curves of the central periodic orbit and of the unstable orbit of period-8 cannot intersect because between them there are invariant curves with $0<RN<1/8$ that separate the domains around the central UPO and the one close to $RN=1/8$.

The homoclinic intersections of the asymptotic curves of the main UPO at ($x=-0.519$, $\dot{x}=0$) are shown in \figref{asympts1}a. The asymptotic curves are calculated by taking many initial conditions with very small intervals along the stable and unstable lines found by diagonalizing the `monodromy matrix' of the periodic orbit (20000 initial conditions along intervals of length $10^{-6}$ \cite{contopoulos2002order}). After 13 intersections the unstable asymptotic curves (orange upwards and burgundy downwards) and the stable curves (cyan upwards and green downwards) reach the neighborhood of two period-1 orbits at ($x = 0.754$, $\dot{x} = \pm 0.944$) where we see the first homoclinic intersections between the orange and cyan curves and between the burgundy and green curves. In the same figure (\figref{asympts1}a) we have also drawn the initial parts of the asymptotic curves of two period-1 orbits. We see the first heteroclinic intersections of these curves with some asymptotic curves of the central periodic orbit. Note that the intersections are between a stable and an unstable curve each time, while the stable (unstable) curves cannot intersect other stable (unstable) curves. Note further that the various asymptotic curves have a symmetry with respect to the x-axis (orange with green, and burgundy with cyan). If we consider all the asymptotic curves of the period-1 periodic orbits we have \figref{asympts1}b. There we see many oscillations of the asymptotic curves and many heteroclinic intersections around the 3 periodic orbits of period-1.

\begin{figure}[H]
\centering
\includegraphics[scale=0.24]{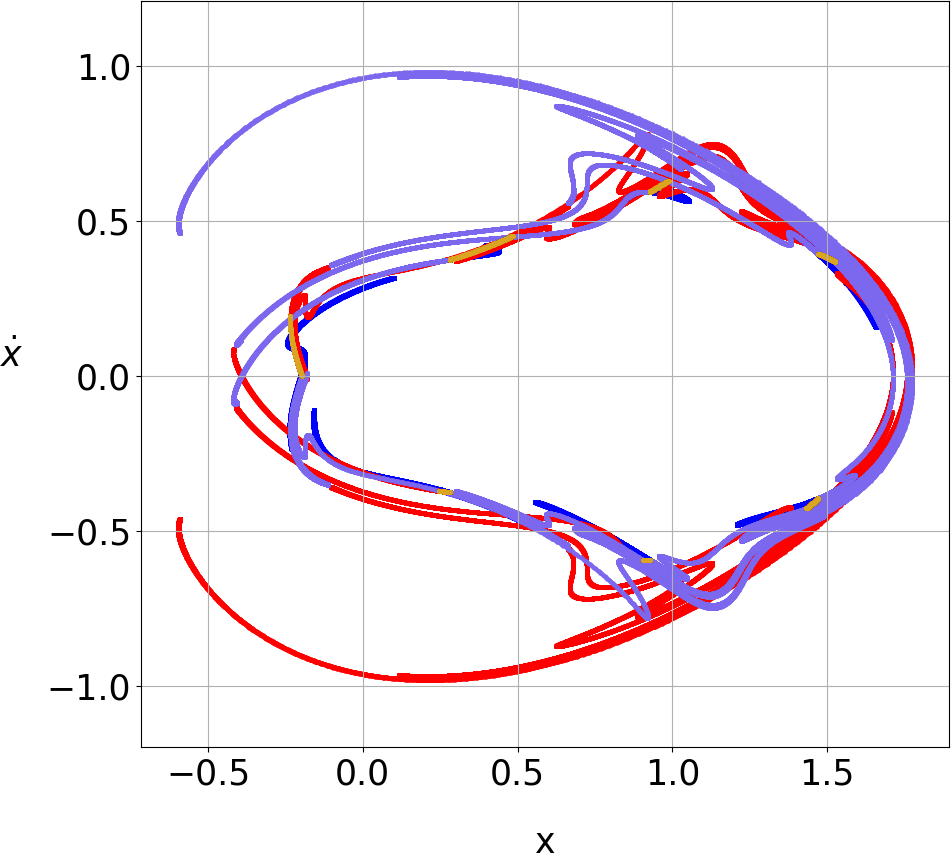}
\includegraphics[scale=0.24]{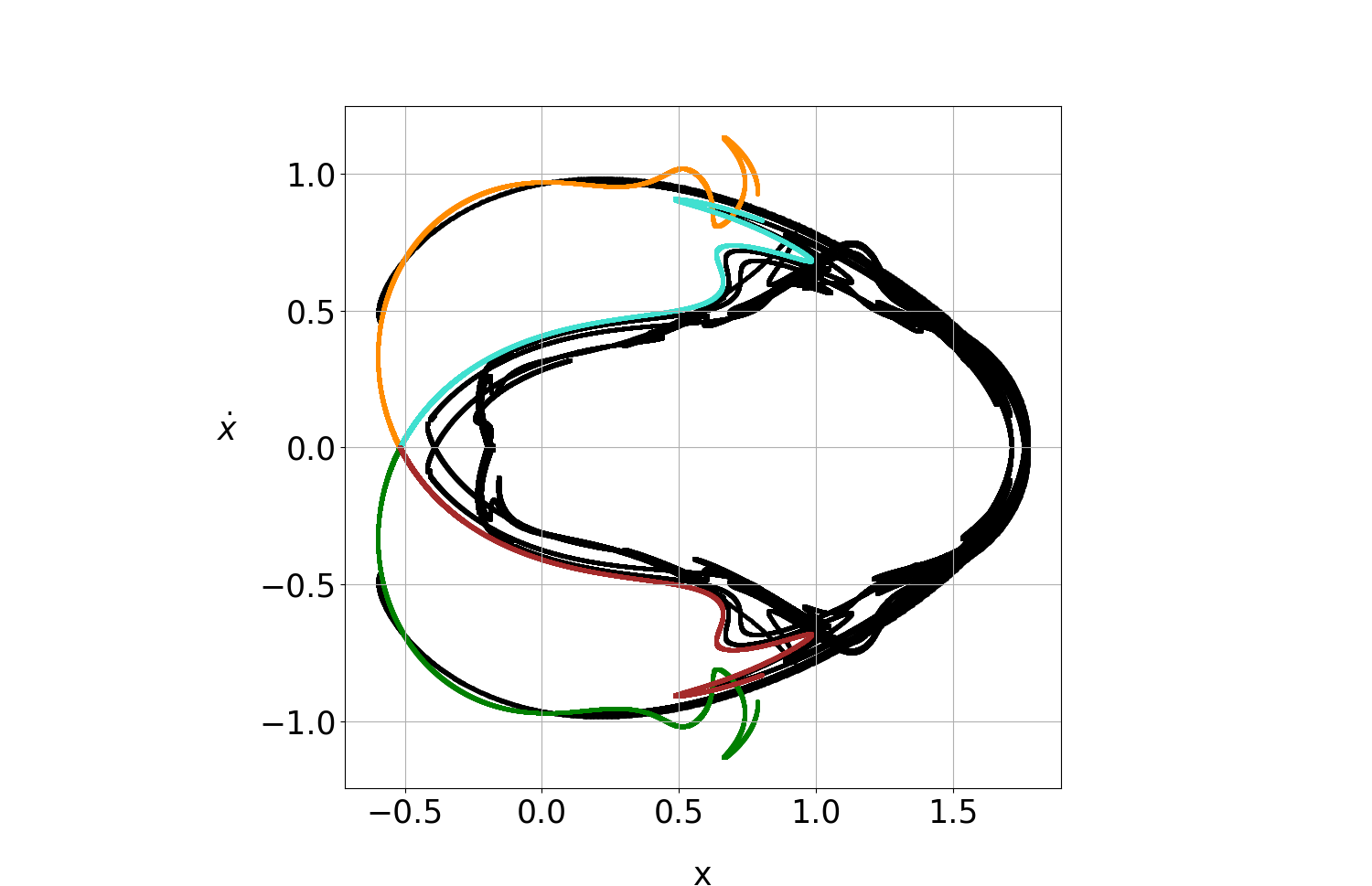}
\caption{(a) Asymptotic curves for $\epsilon = 0.35$, $\alpha = -1/3$ of a period-7 UPO on the x-axis at $x_o = -0.197$. We have two asymptotic curves upwards from the initial point (unstable red and stable purple), and from the other 6 points, up to a total number of 42 intersections (6 images of the initial segment for every of the 7 points). The red and purple curves intersect at many homoclinic points. (b) The asymptotic curves of the central UPO (13 iterations, 4 colors) and of the period-7 orbit (42 iterations, black color).}\label{fig:asympts2}
\end{figure}

If we consider now another UPO in the large chaotic region of \figref{minus1over3}b, we have further homoclinic and heteroclinic intersections. For example in \figref{asympts2}a, we see a period-7 orbit and two asymptotic curves (stable purple and unstable red) from each of the 7-points. Near every point we have 6 iterations of a small initial segment along the stable and unstable directions of each point of the periodic orbit. Thus, the total number of intersections is 42 ($6\times 7$). We see many homoclinic intersections of the red and purple  curves. Then in \figref{asympts2}b we have the asymptotic curves of the UPOs with periods 1 and 7. These intersect at many heteroclinic points.

\begin{figure}[H]
\centering
\includegraphics[scale=0.25]{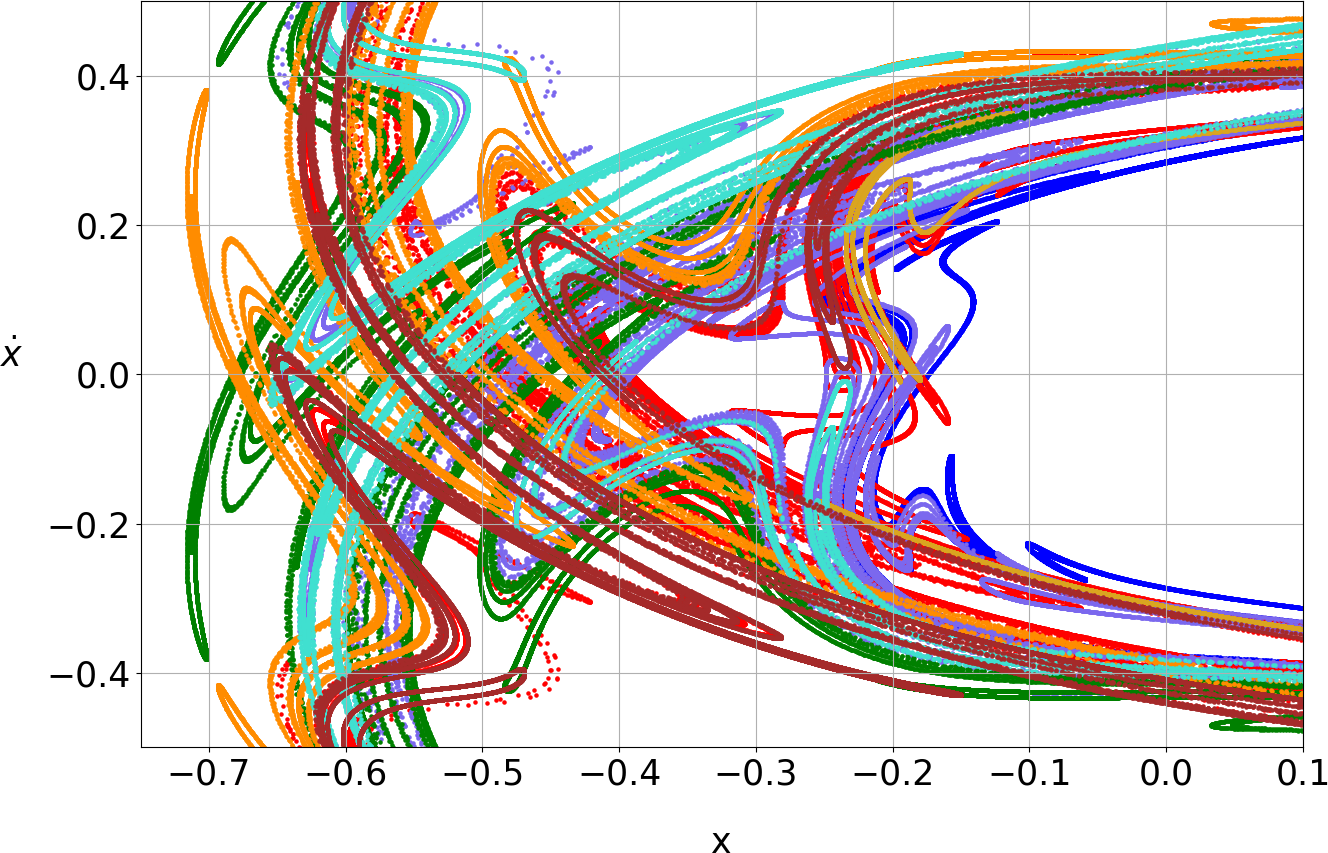}
\caption{A zoom of several asymptotic curves of the central period-1 orbit (20 intersections) and of the period-7 orbit (49 intersections). This is a typical case of resonance overlap which generates chaos.}\label{fig:asympts3}
\end{figure}

A zoom of more intersections between the asymptotic curves of the UPOs of periods 1 and 7 is given in (\figref{asympts3}). We see many homoclinic and heteroclinic intersections. This picture gives a clear indication of chaos.

\section{Conclusions}

In the present paper we applied the theory of the third integral in various potentials of the form
\begin{equation}
V=\frac{1}{2}(\dot{x}^2+x^2+\dot{y}^2+y^2)+\epsilon(xy^2+\alpha x^3),
\end{equation}
for various values of $\epsilon$ and $\alpha$, mainly in cases where we do not have escapes to infinity. We use this integral by calculating the invariant curves on the Poincare surface of section $(\dot{x},x)$ at $y=0$. We used a truncated form of the integral up to order $\epsilon^2$ and compared our results with the exact surfaces of section produced by calculating numerically many orbits.

As it is well known, the system is integrable if we set $\alpha=1/3$. Thus we study the changes found by changing gradually $\alpha$ from $-1/3$ to $1/3$, but we calculate also cases with $\alpha<-1/3$ and $\alpha>1/3$.

In all cases with  $\alpha\neq 1/3$ the system is non integrable and has some chaos, if the perturbation $\epsilon$ is larger than zero. However, for small $\epsilon$ the chaotic domains are very small, and only if the perturbation $\epsilon$ is larger than a critical value $\epsilon_{crit}$ for $\alpha < 0$, which is of the order of the escape perturbation $\epsilon_{esc}$ the chaotic domains increase abruptly. Chaos emerges due to resonance overlap, when the asymptotic curves of different UPO intersect each other.

The main results of our paper are the following:

\begin{enumerate}
    \item We found formal integrals of motion for the H\'{e}non-Heiles and similar Hamiltonians, and then we used them up to order $\epsilon^2$ in calculating theoretical invariant curves of the Poincar\'{e} surfaces of section.
    \item This Hamiltonian is resonant because the frequencies along $x$ and $y$  directions are equal ($\omega_x = \omega_y = 1$). Then if we start calculating the successive terms of an integral $\Phi = \Phi_0 + \epsilon\,\Phi_1 + \epsilon^2\,\Phi_2...$, we find secular terms of order $\epsilon^2$. The secular terms can be eliminated if instead of $\Phi_{1,0} = \frac{1}{2}\left(x^2+\dot{x}^2\right)$, (or $\Phi_{2,0} = \frac{1}{2}\left(y^2+\dot{y}^2\right)$), we use a particular form of the zero order integral $\phi_0=C_0+c_1(2\phi_1)^2+c_2(2\phi_{10})(2\phi_{20})+c_3(2\phi_{20})^2$, where $c_1, c_2, c_3$ are appropriate constants and $C_0=(2\phi_{10})(2\phi_{20})\sin(2t_0)$ with $t_0$ an initial time.
    \item We compared the theoretical invariant curves with the corresponding numerical  (exact) invariant curves. For small values of $\epsilon$ the two types of invariant curves are close to each other but for larger values of $\epsilon$ there are deviations, the most important being the appearance of chaos in the exact surfaces of section around the UPOs.  If $\epsilon$ is small then chaos is very limited, but for $\epsilon$ approaching the escape value $\epsilon_{esc}$ for $\alpha\leq 0$ chaos increases abruptly. For $\epsilon > \epsilon_{\text{esc}}$ most orbits escape to infinity.
    \item In the cases with $\alpha>0$ there are escapes to $\pm \infty$ along the $y$ direction and the escape value $\epsilon = \epsilon_\text{esc}^\prime$ is smaller than $\epsilon_\text{esc}$. E.g. for $\alpha = 0.3$ we have $ \epsilon_\text{esc}^\prime = 0.296$ while $ \epsilon_\text{esc} = 0.454$. Therefore in these cases we have escapes along the y direction as $\epsilon$ increases before the escapes towards $x \rightarrow -\infty$. As $\epsilon$ increases, the domains of escaping orbits along the $y$-axis increase and tend to occupy most of the surface of section, except for an island of stability near the center. Chaos  increases as well as $\epsilon$ increases but remains limited even when $\epsilon$ tends to $\epsilon_\text{esc}$.
    \item We studied in particular the case $\alpha = 1/3$, which corresponds to an integrable  Hamiltonian system. There we have another exact integral besides the total energy, and the invariant curves can be calculated exactly. If we try to apply the general theory of resonant Hamiltonians in this case, we find agreement on the forms of the invariant curves if we truncate the formal integral at order $\epsilon$, but disagreement if we include terms of order $\mathcal{O}(\epsilon^2)$ or higher.
    \item We also studied the case $\alpha = 0$ with  large $\epsilon$ ($\epsilon = 0.3$ and $\epsilon=0.35$) when chaos begins to become important. 
    \item We found that if the Hamiltonian has also further terms of order $\epsilon$  besides $\epsilon(xy^2+ax^3)$, of the form $\epsilon\left(\beta x^2 y + \gamma y^3\right)$, we  cannot eliminate all the secular terms of order $\epsilon^2$.
    \item We studied the onset of chaos in the particular case $\alpha = -1/3$ (the original H\'{e}non-Heiles case). Chaos appears near the UPOs. One way to locate there orbits along the x-axis is by calculating the rotation number ($RN$) of orbits along invariant curves of successive distances $x-x_o$ from a central stable periodic orbit on the right of the origin, for various values of $\epsilon$. For every rational value of $RN$, there is a corresponding periodic orbit at $x = x_\text{per}$. If this is stable, the rotation curve has a small plateau around the value $x=x_\text{per}$ of the periodic orbit, while if the orbit is unstable, the value of $RN$ varies abruptly in a small interval $\Delta x$ around the periodic orbit.
    \item In the above interval $\Delta x$, there are many high order rational values of $RN$ that correspond to unstable orbits. Their asymptotic curves intersect those of the unstable orbit $x=x_\text{per}$. Thus, we have resonance overlap and chaos. Chaos is important (i.e. it affects a large region in phase space) if we have resonance overlap of the asymptotic curves of distant UPOs. The intersections of the asymptotic curves of the same orbit are homoclinic points while the intersections of asymptotic curves of different periodic orbits are heteroclinic points. We gave some numerical examples of the asymptotic curves of some periodic orbits that show these homoclinic and heteroclinic intersections and generate chaos.
\end{enumerate}

\section*{Acknowledgements}
This research was conducted in the framework of the 
program of the Research Committee of the 
 Academy of Athens “Study of order and chaos in 
 quantum dynamical systems.” (No. 200/1026).

\section*{References}
\bibliographystyle{iopart-num}
\bibliography{bibliography}

\end{document}